\documentclass[onefignum, onetabnum]{siamonline220329}

\usepackage{lipsum}
\usepackage{amsfonts}
\usepackage{graphicx}
\usepackage{epstopdf}
\usepackage{algorithmic}
\ifpdf
  \DeclareGraphicsExtensions{.eps,.pdf,.png,.jpg}
\else
  \DeclareGraphicsExtensions{.eps}
\fi

\usepackage{enumitem}
\setlist[enumerate]{leftmargin=.5in}
\setlist[itemize]{leftmargin=.5in}

\newsiamremark{remark}{Remark}
\newsiamremark{hypothesis}{Hypothesis}
\crefname{hypothesis}{Hypothesis}{Hypotheses}
\newsiamthm{claim}{Claim}

\headers{RDA-INR}{Sven Dummer, Nicola Strisciuglio, Christoph Brune}

\title{RDA-INR: Riemannian Diffeomorphic Autoencoding via Implicit Neural Representations}

\author{\href{https://orcid.org/0000-0003-0145-5069}{\includegraphics[scale=0.06]{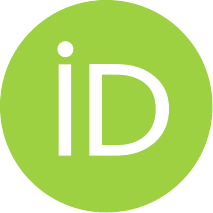}\hspace{1mm} Sven Dummer} \thanks{Mathematics of Imaging \& AI (MIA), Department of Applied Mathematics, University of Twente, Drienerlolaan 5, Enschede 7522~NB,the Netherlands (\email{s.c.dummer@utwente.nl}, \email{c.brune@utwente.nl})}
\and
\href{https://orcid.org/0000-0002-7478-3509}{\includegraphics[scale=0.06]{orcid.pdf}\hspace{1mm} Nicola Strisciuglio} \thanks{Data Management \& Biometrics (DMB), Department of Computer Science, University of Twente, Drienerlolaan 5, Enschede 7522~NB, the Netherlands (\email{n.strisciuglio@utwente.nl})}
\and
\href{https://orcid.org/0000-0003-0145-5069}{\includegraphics[scale=0.06]{orcid.pdf}\hspace{1mm} Christoph Brune} \footnotemark[1] 
}

\usepackage{amsopn}

\newcommand\norm[1]{\left\lVert#1\right\rVert}
\DeclareMathOperator*{\expectation}{\mathop{\mathbb{E}}}
\DeclareMathOperator*{\argmin}{arg\,min}

\usepackage{utfsym}
\newcommand{\cross}{\scalebox{0.75}{\usym{2613}}}

\usepackage{booktabs}       % professional-quality tables
\usepackage{multirow}
\usepackage{float}
\usepackage[caption=false]{subfig}

\ifpdf
\hypersetup{
  pdftitle={RDA-INR: Riemannian Diffeomorphic Autoencoding via Implicit Neural Representations},
  pdfauthor={Sven Dummer, Nicola Strisciuglio, Christoph Brune}
}
\fi

\usepackage{xcolor}

\definecolor{newcolor}{rgb}{.8,.349,.1}

\begin{document}

\maketitle

% REQUIRED
\begin{abstract}
Diffeomorphic registration frameworks such as Large Deformation Diffeomorphic Metric Mapping (LDDMM) are used in computer graphics and the medical domain for atlas building, statistical latent modeling, and pairwise and groupwise registration. In recent years, researchers have developed neural network-based approaches regarding diffeomorphic registration to improve the accuracy and computational efficiency of traditional methods. In this work, we focus on a limitation of neural network-based atlas building and statistical latent modeling methods, namely that they either are (i) resolution dependent or (ii) disregard any data- or problem-specific geometry needed for proper mean-variance analysis. In particular, we overcome this limitation by designing a novel encoder based on resolution-independent implicit neural representations. The encoder achieves resolution invariance for LDDMM-based statistical latent modeling. Additionally, the encoder adds LDDMM Riemannian geometry to resolution-independent deep learning models for statistical latent modeling.  We investigate how the Riemannian geometry improves latent modeling and is required for a proper mean-variance analysis. To highlight the benefit of resolution independence for LDDMM-based data variability modeling, we show that our approach outperforms current neural network-based LDDMM latent code models. Our work paves the way for more research into how Riemannian geometry, shape respectively image analysis, and deep learning can be combined.
\end{abstract}

% REQUIRED

% REQUIRED

\section{Introduction}
Shape and image registration are important tools for analyzing object similarities and differences. To register two images $I$ and $J$, a function $f(x)$ is constructed such that point $f(x)$ in image $J$ corresponds to $x$ in image $I$. For instance, when comparing two human brain images, one wants to match points on the same subpart of the brain. When instead matching two shapes $\mathcal{S}$ and $\mathcal{T}$, a function $f(x)$ should assign to each point $x$ on $\mathcal{S}$ a point $f(x)$ on $\mathcal{T}$ in a meaningful way. For example, when comparing two human shapes, one wants to match points on the arm of one human with points on the same arm of the other human. These image and shape registrations have, for instance, applications in the medical domain. As the human anatomy varies from person to person, registration algorithms transform these anatomies into a single coordinate frame. Once the anatomies are registered, it is possible to compare them and establish correspondences between them. 

Various diffeomorphic registration frameworks are available in the literature, for instance, Stationary Velocity fields (SVF) \cite{arsigny2006log, ashburner2007fast, vercauteren2009diffeomorphic} or Large Deformation Diffeomorphic Metric Mapping (LDDMM). One specific registration task is pairwise registration where a registration between exactly two objects is calculated \cite{beg2005computing, vaillant2005surface, miller2006geodesic, vialard2012diffeomorphic, Younes2019, zhang2019fast, amor2022resnet}. To register multiple objects simultaneously, groupwise registration algorithms \cite{joshi2004unbiased, ma2008bayesian, ma2010bayesian, Younes2019} register all the objects to the same template. In this process, the template can also be estimated to perform atlas building. The estimated atlas is used for proper assessment of the differences between objects. Finally, besides plain diffeomorphic registration, LDDMM is also used for modeling the variability of the data. Specifically, using the LDDMM Riemannian distance, one can use the manifold variant of principal component analysis (PCA), called principal geodesic analysis (PGA), to estimate the factors of variation given the data \cite{zhang2015bayesian, zhang2017probabilistic}. The Lagrangian nature of LDDMM acts as a physical modeling basis (physical consistency) providing access to Fréchet mean and variance. More precisely, we use the term physical consistency as LDDMM's geodesic structure enables time-dependent non-autonomous behavior and is reminiscent of Lagrangian structures used in physics via the least-action principle. 

Recently, the different registration tasks have also been addressed via neural network methods. Table \ref{tab:methods_vs_properties}, positioned near the end of the introduction, presents and compares a comprehensive selection of them. Specifically, there are neural network models that speed up the computation of pairwise registrations \cite{joshi2022diffeomorphic, krebs2019learning, shen2019networks, yang2017quicksilver, jia2023fourier, wang2020deepflash}, groupwise registrations \cite{yang2017quicksilver, dalca2019learning, ding2022aladdin, mok2020large, wang2022geo}, and joint groupwise registration and atlas building \cite{dalca2019learning, ding2022aladdin, wang2022geo}. Furthermore, attempts have been made to improve LDDMM PGA by adding neural networks \cite{bone2019learning, hinkle2018diffeomorphic}. Finally, all the approaches described above require a specific \textit{spatial discretization} of the objects and the underlying registration domain. Often one does not a-priori know the optimal discretization, one might get memory issues when one refines the discretization, and neural network models might not generalize well to other resolutions. Recently, resolution-independent methods avoiding this \textit{spatial discretization} showed improved performance for diffeomorphic registration \cite{sun2022topology, zou2023homeomorphic, wu2023neurepdiff, han2023diffeomorphic, tian2023texttt} and non-diffeomorphic registration \cite{wolterink2022implicit}. This motivates the need for having resolution-independent methods for LDDMM PGA, which achieves joint atlas building and statistical latent modeling.  

In the literature, there are two types of approaches for joint atlas building and statistical latent modeling. On the one hand, we have the aforementioned LDDMM literature that considers methods that have LDDMM physical consistency \cite{bone2019learning, hinkle2018diffeomorphic} but are not resolution-independent. On the other hand, we have models that consider resolution-independent implicit neural representation (INR) methods for shape data \cite{deng2021deformed, zheng2021deep, sun2022topology} but do not consider LDDMM physical consistency. This physical consistency is required for properly dealing with longitudinal data and to properly calculate the Fréchet mean and variance of the dataset:
\begin{equation*}
    \mu = \argmin_O \frac{1}{N}\sum_{i=1}^N d(O, O_i)^2, \quad \sigma^2 = \frac{1}{N}\sum_{i=1}^N d(\mu, O_i)^2
\end{equation*}
where $d$ is the Riemannian distance induced by LDDMM and $\{O_i\}_{i=1}^N$ are $N$ images or shapes. When the model does not consider this Riemannian structure, one might not calculate the actual geodesics, which are needed for calculating a proper estimate of the variance. Moreover, the estimated mean might not be consistent with the data, which is detrimental if you want to study a dataset via a sensible average or atlas. 

\subsection{Contributions}
\begin{figure}[b!]
    \centering
    \includegraphics[width=0.7\textwidth]{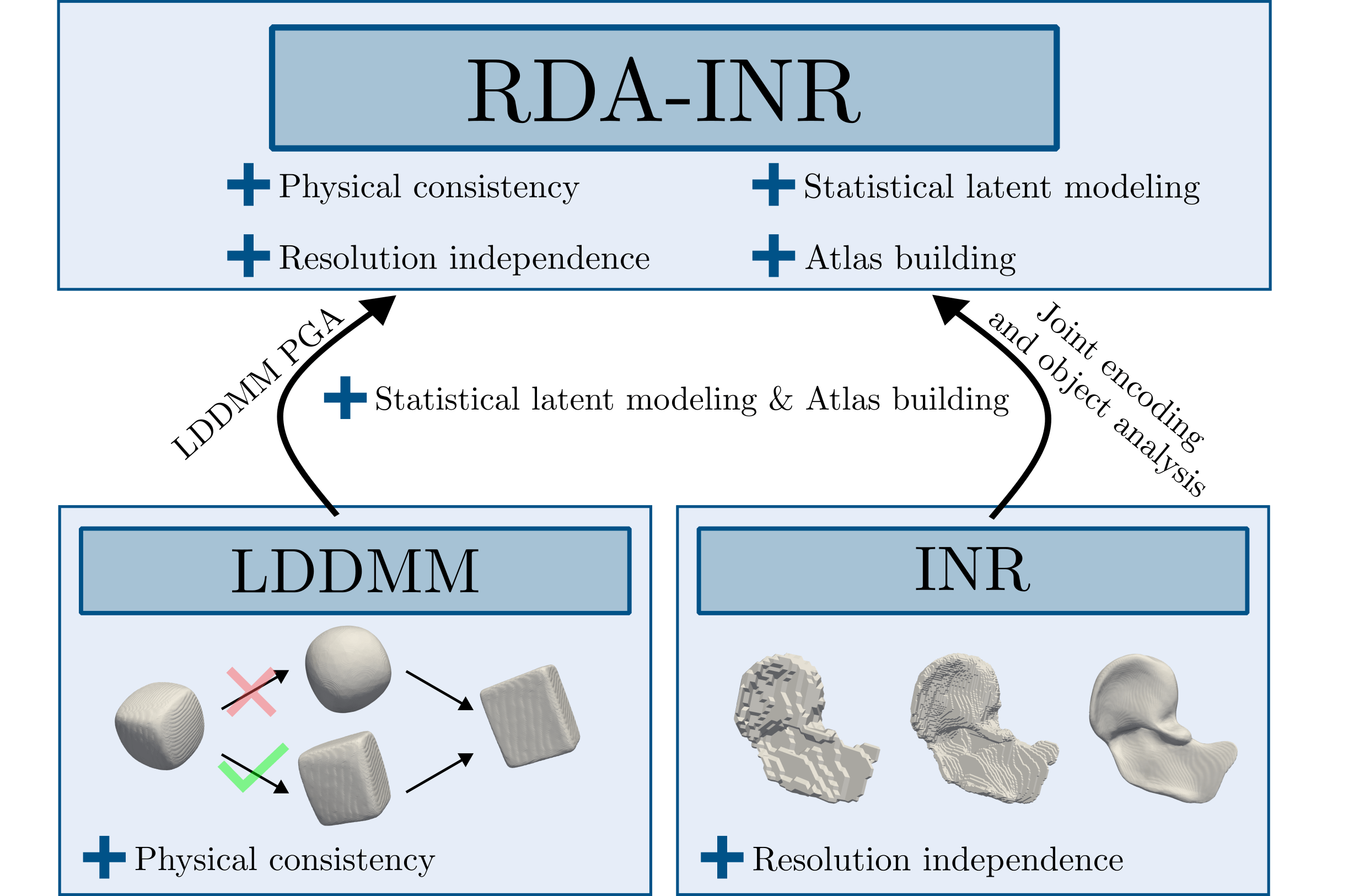}
    \caption{\textbf{Overview of our RDA-INR framework.} We combine LDDMM PGA and resolution-independent implicit neural representation (INR) methods used for joint encoding and registration. This combination yields a method for statistical latent modeling and atlas building that is (i) physically consistent via LDDMM Riemannian geometry and (ii) resolution independent. \label{fig:story_of_paper}}
\end{figure}

\begin{figure}[t!]
    \centering
    \includegraphics[width=0.75\textwidth]{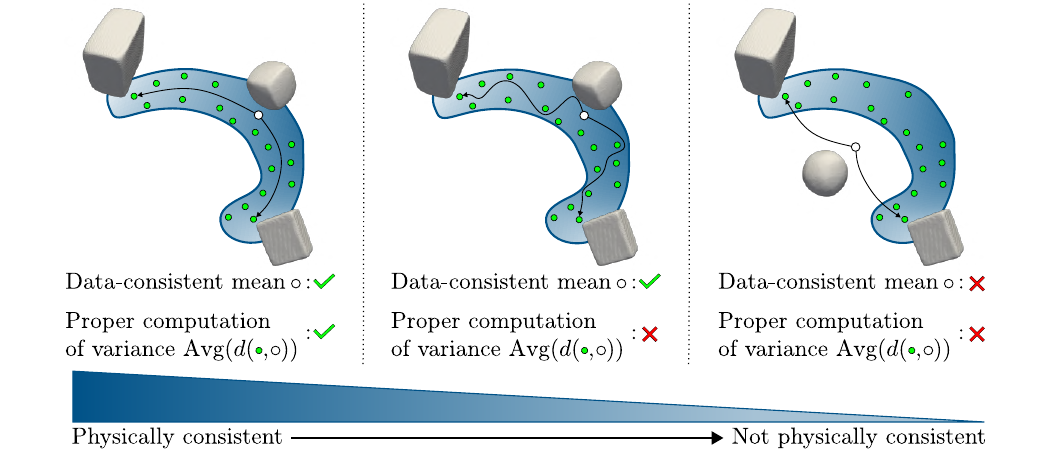}
    \caption{\textbf{Latent model classes for mean-variance analysis.} The calculation of the Fréchet mean and the variance of a dataset depend on the length of the geodesics between objects. However, not all latent models take this into account, which results in three categories of methods: methods considering geodesics and obtaining a data-consistent Fréchet mean, methods that do not consider geodesics but still obtain a 
    Fréchet mean consistent with the data, and methods that do not obtain such a Fréchet mean.\label{fig:possible_problem_with_mean_and_variance_calculation}}
\end{figure}

Our work addresses the previously mentioned issues of joint atlas building and statistical latent modeling to study the benefits of (i) LDDMM physical consistency and (ii) resolution-independence properties. Specifically, we connect the shape INR literature with the LDDMM PGA literature, where the latter focuses on image data. To have a general viewpoint, we address this connection mainly from a point cloud or mesh perspective instead of a segmentation mask perspective. Our main contributions are:
\begin{enumerate}
    \item We introduce a novel model called Riemannian Diffeomorphic Autoencoding via Implicit Neural Representations (RDA-INR) that deals with the aforementioned issues. Figure \ref{fig:story_of_paper} shows how our model is designed. Specifically, our RDA-INR adds:
    \begin{itemize}
        \item Riemannian geometry to the INR literature on shape variability modeling by using LDDMM PGA and its Riemannian regularization functional,
        \item resolution-independent INRs to LDDMM PGA.
    \end{itemize}
    \item Our latent model is the first in the registration literature using image-based (implicit neural) data representations for dealing directly with point cloud or mesh data. 
    \item We show that the Riemannian regularization functional is required for a proper mean-variance analysis, i.e.\ avoiding the middle and right category in Figure \ref{fig:possible_problem_with_mean_and_variance_calculation}. More precisely, the Riemannian geometry is needed to simultaneously obtain: a template as Fréchet mean of the data and physically plausible deformations aligning the template to the data. 
    \item We show that the Riemannian regularization functional can improve the latent modeling compared to a non-Riemannian regularization functional. More precisely, it can lead to improved reconstruction generalization and more robustness to noise. 
    \item By comparing our model to a mesh-based neural network latent model from the LDDMM literature, we show that resolution independence yields higher-quality templates and higher-quality reconstructions.
\end{enumerate}
\begin{table}[t!]
    \footnotesize
    \caption{\textbf{Neural network-based diffeomorphic registration methods.} The columns show the features of the papers: atlas building, statistical latent modeling, physical consistency using LDDMM Riemannian geometry, and resolution invariance. Our novel RDA-INR framework is the only work that addresses all the features.}
	\centering
	\begin{tabular}{c||c|c|c|c}
		\toprule
		\multicolumn{1}{c@{\quad}}{\textbf{\textit{Model}}} & \multicolumn{4}{c}{\textbf{\textit{Features}}}                 \\
		\cmidrule(r){1-1} \cmidrule{2-5}
 &  Atlas building & Stat. lat. modeling& Riemannian & Res. indep.\\
\cmidrule{1-5} \morecmidrules \cmidrule{1-5}
  ResNet-LDDMM \cite{amor2022resnet} & \cross & \cross & \checkmark & \cross \\
  R2Net \cite{joshi2022diffeomorphic} & \cross & \cross & \cross & \cross \\
  Krebs et al.\ \cite{krebs2019learning} & \cross & \checkmark & \cross & \cross \\
  Shen et al.\ \cite{shen2019networks} & \cross & \cross & \cross & \cross \\
  Quicksilver \cite{yang2017quicksilver} & \cross & \cross & \checkmark & \cross  \\ 
  Niethammer et al.\ \cite{niethammer2019metric} & \cross & \cross & \cross & \cross  \\ 
  Diff.~Voxelmorph \cite{dalca2019learning} & \checkmark & \cross & \cross & \cross \\ 
  DNVF \cite{han2023diffeomorphic} & \cross & \cross & \cross & \checkmark \\ 
  NePhi \cite{tian2023texttt} & \cross & \cross & \cross & \checkmark \\ 
  Zou et al.\ \cite{zou2023homeomorphic} & \cross & \cross & \cross & \checkmark \\ 
  Fourier-Net+ \cite{jia2023fourier} & \cross & \cross & \cross & \cross  \\ 
  DeepFLASH \cite{wang2020deepflash} & \cross & \cross & \checkmark & \cross \\ 
  LapIRN \cite{mok2020large} & \cross & \cross & \cross & \cross \\ 
  Aladdin \cite{ding2022aladdin} & \checkmark & \cross & \cross & \cross \\
  Hinkle et al.\ \cite{hinkle2018diffeomorphic}  & \checkmark & \checkmark & \checkmark & \cross  \\
  DAE \cite{bone2019learning} & \checkmark & \checkmark & \checkmark & \cross  \\
  Geo-Sic \cite{wang2022geo} & \checkmark & \cross & \checkmark & \cross \\
  NeurEPDiff \cite{wu2023neurepdiff}  & \cross & \cross & \checkmark & \checkmark \\
  FlowSSM \cite{ludke2022landmark} & \cross & \checkmark & \cross & \checkmark \\
  NMF \cite{gupta2020neural} & \cross & \checkmark & \cross & \checkmark \\
  NDF \cite{sun2022topology} & \checkmark & \checkmark & \cross & \checkmark \\
  RDA-INR (Ours) & \checkmark & \checkmark & \checkmark & \checkmark \\ 
		\bottomrule
	\end{tabular}
    \label{tab:methods_vs_properties}
\end{table}

\subsection{Outline}
We organize the rest of the paper as follows. In Section \ref{sec:related_works}, we discuss the related works. Subsequently, in Section \ref{sec:preliminaries}, we treat the background theory needed for our method. Specifically, we treat INRs and the implicit representation of shapes (Section \ref{sec:Impl_Shape_Representation}), the LDDMM framework (Section \ref{sec:LDDMM}), and LDDMM PGA (Section \ref{sec:LDDMM_PGA}). Section \ref{sec:our_model} introduces our neural network model. In Section \ref{sec:numerical_results} we compare this model to the same model with a non-Riemannian regularization. We also compare our method to DAE \cite{bone2019learning}. Finally, we summarize the paper, discuss some limitations, and discuss future work in Section \ref{sec:conclusion}. In the appendices, we provide additional experiments and details, and some basics of Riemannian geometry. Specifically, the appendices present an interesting result: our Riemannian regularization functional is novel in the LDDMM literature, yet still related to the standard LDDMM regularizers.

\section{Related work} \label{sec:related_works}
In this section, we contextualize our contributions by discussing related literature on principal geodesic analysis (PGA), Riemannian geometry for latent space models, and neural ordinary differential equations (NODE). 

\subsection{Principal geodesic analysis}
Principal component analysis (PCA) is a well-known technique that has applications in, among others, dimensionality reduction, data visualization, and feature extraction. In addition, the probabilistic extension of PCA, called Probabilistic PCA, can also be used to generate new samples from the data distribution \cite{bishop2006pattern}. A limitation of PCA and probabilistic PCA is that they can only be applied to Euclidean data and not Riemannian manifold-valued data. To solve this limitation, principal geodesic analysis (PGA) \cite{fletcher2004principal} and probabilistic principal geodesic analysis (PPGA) \cite{zhang2013probabilistic} are introduced as extensions of PCA and probabilistic PCA, respectively. 

PGA is applied in various domains. For instance, optimal transport (OT) uses PGA to define Fréchet means and interpolations of data distributions called barycenters \cite{agueh2011barycenters, seguy2015principal}. Shape analysis uses PGA as well \cite{fletcher2004principal}. Different versions of shape PGA are obtained by choosing different Riemannian distances on the shape space. For instance, while in Zhang and Fletcher \cite{zhang2015bayesian} and Zhang et al.\ \cite{zhang2017probabilistic} an LDDMM-based distance is used for PGA, Heeren et al.\ \cite{heeren2018principal} use a thin shell energy as the Riemannian distance. Furthermore, besides shape analysis based on PGA, shape analysis tools exist that use tangent PCA \cite{vaillant2004statistics, Younes2019} or that use a combination of tangent PCA and geodesic PCA \cite{charlier2018distortion}. 

Similar to how an autoencoder \cite{kramer1991nonlinear, hinton2006reducing} is a nonlinear version of PCA, our encoding framework can be viewed as a nonlinear version of PGA and PPGA. Another neural network model that can be regarded as a nonlinear version of PPGA is the Riemannian variational autoencoder (Riemannian VAE) \cite{miolane2020learning}. The Riemannian VAE only considers learning a latent space for data on finite-dimensional Riemannian manifolds for which the exponential map and the Riemannian distance have an explicit formula. Hence, it can not be used for the infinite-dimensional Riemannian manifold of shapes and images for which the LDDMM Riemannian distance does not have an explicit formula. Our deformable template model extends the Riemannian VAE to this setting.

Bone et al.\ \cite{bone2019learning} and Hinkle et al.\ \cite{hinkle2018diffeomorphic} construct an autoencoder that resembles LDDMM PGA \cite{zhang2015bayesian}. While Hinkle et al.\ \cite{hinkle2018diffeomorphic} is only applicable to images, Bone et al.\ \cite{bone2019learning} is also applicable to meshes. Similar to our work, both use an ordinary differential equation (ODE) to deform a learned template into a reconstruction. Moreover, they introduce a regularizer on the velocity vector fields to let the ODE flow be a diffeomorphism.
% and to make the learned template a Fréchet mean of the shape mesh data. 
In contrast, while they use a specific resolution for the images, shapes, and diffeomorphisms, we construct a resolution-independent method. For instance, while Bone et al.\ \cite{bone2019learning} consider the mesh representation of shapes, we consider the state-of-the-art implicit shape representation. Combining this representation with INRs allows us to acquire a resolution-independent approach where we obtain a mesh or point cloud at an arbitrary resolution by applying marching cubes \cite{lorensen1987marching}.

\subsection{Riemannian geometry for latent space models} \label{sec:rel_work_riem_geom} 
Latent space models based on neural networks are used for, among others, dimensionality reduction, clustering, and data augmentation. Most of these models have the drawback that distances between latent codes do not always correspond to a 'semantic' distance between the objects they represent. One mathematical tool to solve this issue is Riemannian geometry. Arvanitidis et al.\ \cite{arvanitidis2017latent, arvanitidis2021prior} design a Riemannian distance between latent codes by defining a Riemannian metric. This metric is the standard Euclidean metric pulled back under the action of the decoder or a surrogate metric approximating the pullback. Shortest paths under these Riemannian metrics are found by solving the geodesic equation \cite{arvanitidis2017latent, arvanitidis2019fast} or the geodesic minimization problem \cite{chen2018metrics, chadebec2020geometry}. The resulting paths yield good distances and create good interpolations. Furthermore, the Riemannian metric can be used for improved clustering \cite{yang2018geodesic, chadebec2020geometry} and for improved generative modeling. In particular, Chadebec et al.\ \cite{chadebec2020geometry} incorporate the Riemannian metric into a Hamiltonian Markov Chain that samples from a variational autoencoder's posterior distribution. 

Our novel latent model that uses implicit neural representations (INRs) is related to the Riemannian geometry for latent space models literature. Although the Euclidean distances in our latent space do not necessarily approximate LDDMM distances, we do embed a Riemannian distance between a learned template and the other reconstructed data points by using LDDMM as the Riemannian distance. The most relevant work that also attempts to incorporate shape Riemannian geometry into an INR latent space model is Atzmon et al.\ \cite{atzmon2021augmenting}. However, their model can not perform shape encoding and shape registration jointly. Moreover, it does not enforce a diffeomorphic structure. 

\subsection{Neural dynamics}
The neural ODE (NODE) \cite{chen2018neural} is a first-order differential equation where the time derivative is parameterized by a neural network. It has applications in learning dynamics \cite{greydanus2019hamiltonian, cranmer2020lagrangian, tong2020trajectorynet}, control \cite{onken2022neural, wotte2022discovering}, generative modeling \cite{chen2018neural, yang2020potential}, and joint shape encoding, reconstruction, and registration \cite{niemeyer2019occupancy, gupta2020neural, sun2022topology}. Furthermore, there is a bidirectional connection between the NODE and optimal transport (OT). First, due to the fluid dynamics formulation of OT \cite{benamou2000computational}, NODEs can be used to solve high-dimensional OT problems \cite{ruthotto2020machine}. In addition, the learned dynamics can be complex and require many time steps to be solved accurately. As a consequence, training NODEs can be challenging and time-consuming. OT-inspired regularization functionals solve this issue as they simplify the dynamics and reduce the time needed to solve the NODE \cite{finlay2020train, onken2021ot}.

Our framework fits the bidirectional connection of NODEs and OT. First, we use a neural ODE to solve the LDDMM problem, which is a problem that in formulation bears similarities to the fluid dynamics formulation of OT. Hence, our approach shares similarities with the works using NODEs to solve OT problems. In addition, we use cost functionals inspired by LDDMM to regularize the NODE used for reconstruction and registration. This regularization fits the works that use OT for regularizing the dynamics of NODEs. 

\section{Preliminaries}
\label{sec:preliminaries}

\subsection{Implicit neural representations and implicit shape representations} \label{sec:Impl_Shape_Representation}
In many variational problems, we optimize for an unknown image, a solution to a partial differential equation, a velocity vector field for registration, or a shape. The idea behind an implicit neural representation (INR) is to represent such objects by a neural network $f_\theta \colon \Omega \rightarrow \mathbb{R}^n$, where $\Omega$ can be a spatial or a spatiotemporal domain. In contrast to, for instance, finite element solutions, which also lie in an infinite dimensional space, the neural network representation avoids a spatial or spatiotemporal discretization of the objects. Furthermore, INRs yield a solution in an infinite dimensional space by optimizing over elements from the finite-dimensional submanifold defined by the neural network weights. Methods employing this approach and using $f_\theta$ as parameterization of a velocity vector field or a deformation field, achieve good performance in diffeomorphic \cite{zou2023homeomorphic, wu2023neurepdiff, han2023diffeomorphic, tian2023texttt} and non-diffeomorphic registration \cite{wolterink2022implicit}. 

When representing shapes with INRs, the implicit shape representation is of interest. A shape $\mathcal{S}:=\{x \mid f(x) = c\}$ is described by an image via the $c$ level set of an image function $f\colon \Omega \rightarrow \mathbb{R}$ on the domain $\Omega\subset \mathbb{R}^d$. Given a shape $\mathcal{S} = \partial \mathcal{S}_{\text{int}}$ with a bounded open set $\mathcal{S}_{\text{int}} \subset \Omega$ as interior, two examples of common implicit representations are the signed distance function (SDF) with $c=0$ and the occupancy function (OCC) with $c=0.5$, respectively:
\begin{align}
    \text{SDF}_\mathcal{S}(x) & =\begin{aligned}
        \begin{cases}
        \inf_{y \in \mathcal{S}}\norm{x - y}_2 \quad \quad \text{if } x  \in \mathcal{S}_{\text{int}}, \\
        -\inf_{y \in \mathcal{S}}\norm{x - y}_2 \quad \text{if } x \notin \mathcal{S}_{\text{int}}, \\
    \end{cases}
    \end{aligned} \label{eq:SDF} \\
    \text{OCC}_\mathcal{S}(x) & = \begin{aligned}
        \begin{cases}
        0 \qquad \qquad \qquad \qquad \text{if } x \in \Omega \backslash \overline{\mathcal{S}_{\text{int}}}, \\
        0.5 \qquad \qquad \qquad \quad \, \text{if } x \in \mathcal{S}, \\
        1 \qquad \qquad \qquad \qquad \text{if } x \in \mathcal{S}_{\text{int}}. \\
    \end{cases}
    \end{aligned} 
    \label{eq:OccFunc}  
\end{align}

Unlike the traditional methods that use point clouds, meshes, or voxel grids, INR methods that use the implicit representation, such as DeepSDF \cite{park2019deepsdf}, SIREN \cite{sitzmann2020implicit}, and Occupancy Network \cite{mescheder2019occupancy}, are spatial discretization-independent. Furthermore, they yield state-of-the-art shape encodings and reconstructions, which are obtained by extracting a level set at arbitrary resolution, for instance, via marching cubes \cite{lorensen1987marching}. However, unlike deforming meshes and point clouds by deforming their vertices, morphing an implicit representation does not immediately create point correspondences between the initial and deformed shapes. To remedy this issue, recent works \cite{deng2021deformed, zheng2021deep, sun2022topology} create template-based INR models that allow for joint registration and encoding. These models parameterize a template shape with an INR and transform it into implicit representations of other shapes via image registration techniques. Subsequently, the model uses the image registrations to the template INR to perform shape matching.

\subsection{Diffeomorphic registration and the LDDMM Riemannian distance} \label{sec:LDDMM}
In this work, we focus on the Large Deformation Diffeomorphic Metric Mapping (LDDMM) registration framework. The objective of LDDMM is to find a diffeomorphism $\phi\colon \Omega \rightarrow \Omega$ that matches two objects $O_1$ and $O_2$ (e.g., images or shapes defined on the spatial domain $\Omega \subset \mathbb{R}^d$) by deforming $O_1$ into $O_2$. More precisely, LDDMM wants $\phi \cdot O_1 = O_2$ with $\phi \cdot O$ a left group action of the diffeomorphism group on the set of objects. This group action formulation is very flexible due to the ambient diffeomorphisms. For instance, a function $I\colon \Omega \rightarrow \mathbb{R}^n$ representing an image or an implicit shape representation can be deformed by a left group action $\phi \cdot I := I \circ \phi^{-1}$. Moreover, a set-based definition of a shape $\mathcal{S}:=\{x \mid x \text{ on the shape or pointcloud}\}$ can be deformed by $\phi \cdot \mathcal{S} := \phi(\mathcal{S})$. 

To construct a diffeomorphism achieving $\phi \cdot O_1 = O_2$, LDDMM considers a subgroup of the diffeomorphism group. This subgroup consists of diffeomorphisms emerging as flows of ordinary differential equations (ODEs). Define a time-dependent vector field $v\colon\Omega \times [0,1] \rightarrow \mathbb{R}^d$ and assume $v(\cdot, t) \in V$ for some Banach space $V$ with norm $\norm{\cdot}_V$. The corresponding ODE is $\frac{\mathrm{d} x}{\mathrm{d} t} = v(x, t)$ and gives us a flow $\phi_t$:
\begin{equation}
    \begin{aligned}
        \frac{\partial \phi_t}{\partial t}(x) & = v(\phi_t(x), t), \quad \phi_0(x) & = x.
    \end{aligned}
    \label{eq:ode_diffeo}
\end{equation}
To make these ODE flows diffeomorphisms, one has to enforce smoothness on the velocity vector fields $v(\cdot, t) \in V$. Generally, this is accomplished by choosing an appropriate Banach space $V$ and assuming $v \in L^1([0, 1], V)$. One class of such spaces is admissible Banach spaces: 

\begin{definition}[Admissible Banach spaces \cite{Younes2019}]
Let $C_0^1(\Omega, \mathbb{R}^d)$ be the Banach space of continuously differentiable vector fields $\nu$ on the open and bounded domain $\Omega \subset \mathbb{R}^d$ such that both $\nu$ and its Jacobian $J\nu$ vanish on $\partial \Omega$. Furthermore, define a Banach space $V \subset C_0^1(\Omega, \mathbb{R}^d)$ and let $\norm{\nu}_{1,\infty}:=\norm{\nu}_\infty + \norm{J\nu}_\infty$ for $\nu \in C_0^1(\Omega, \mathbb{R}^d)$. The Banach space $V$ is admissible if it is (canonically) embedded in $(C_0^1(\Omega, \mathbb{R}^d), \norm{\cdot}_{1,\infty})$. In other words, there exist a constant $C$ such that $\forall \nu \in V$, $\norm{\nu}_V \geq C\norm{\nu}_{1,\infty}$.
\end{definition}

Using these considerations, the subgroup that LDDMM considers is defined as
\vspace*{-5pt}
\begin{equation*}
    G:=\{\phi_1 \mid \phi_t \text{ satisfies Equation \eqref{eq:ode_diffeo} for some } v \in L^1([0, 1], V) \}.
\end{equation*}
There might exist multiple $\phi \in G$ such that $\phi \cdot O_1 = O_2$ for objects $O_1$ and $O_2$. The LDDMM approach selects the 'smallest' $\phi \in G$ such that $\phi \cdot O_1 = O_2$. To define 'smallest', LDDMM uses the following Riemannian distance $d_G$ on $G$:
\begin{equation*}
    \begin{aligned}
        d_G(\phi, \psi)^2 := \inf_{v \in L^2([0,1], V)} \quad & \left( \int_{0}^1 \norm{v(\cdot, t)}_V^2 \mathrm{d}t \right) \\
        \textrm{s.t.} \quad \quad \, \, \, & \frac{\partial \phi_t}{\partial t}(x) = v(\phi_t(x), t), \quad \phi_0(x) = x, \\
        & \psi = \phi_1 \circ \phi.
    \end{aligned}
\end{equation*}
where $\norm{\cdot}_V$ is a norm on $V$. The LDDMM approach selects the $\phi \in G$ such that $\phi \cdot O_1 = O_2$ and $d_G(\text{id}, \phi)$ is small. Alternatively, when $O_1$ and $O_2$ belong to a set $\mathcal{O}$ of diffeomorphic objects, we can reformulate this problem as finding a distance between $O_1$ and $O_2$:
\begin{theorem}[\cite{Younes2019}]
    Assume that $V$ is an admissible Banach space. Then $(G, d_G)$ is a complete metric space. Furthermore, assume we consider a set $\mathcal{O}:=\{\phi \cdot O_{\text{temp}} \mid \phi \in G\}$ for some template object $O_{\text{temp}}$. Then we can define a pseudo-distance $d_\mathcal{O}(O_1, O_2)$ on $\mathcal{O}$ as 
    \begin{equation}
            d_\mathcal{O}(O_1, O_2):=\inf_{\phi} \left( d_G(\normalfont{\text{id}}, \phi) \mid O_2 = \phi \cdot O_1, \phi \in G \right)
            \label{eq:rho_I_v1}
    \end{equation} 
    or alternatively:
    \begin{equation}
        \begin{aligned}
            d_\mathcal{O}(O_1, O_2)^2 = \inf_{v(\cdot, t)} \quad & \left( \int_{0}^1 \norm{v(\cdot, t)}_V^2 \mathrm{d}t \right) \\
            \textrm{s.t.} \quad \, & \frac{\partial \phi_t}{\partial t}(x) = v(\phi_t(x), t), \quad \phi_0(x) = x, \\
            & \phi_1 \cdot O_1 = O_2.
            \label{eq:rho_I_v2}
        \end{aligned}
    \end{equation}
    In addition, $d_\mathcal{O}$ is a distance if the action $\phi \rightarrow \phi \cdot O_{\text{temp}}$ is continuous from $G$ to $\mathcal{O}$ given the topology induced by $d_G$ on $G$ and some topology on $\mathcal{O}$.
\label{thm:object_dist_via_diffeomorphism_group_action}
\end{theorem}

One example when $\phi \rightarrow \phi \cdot O_{\text{temp}}$ is continuous as a mapping from $G$ to $\mathcal{O}$ is when considering the shape $O_{\text{temp}}=\{x\mid f(x)=0\}=\partial A$ for some continuous function $f$ and $A\subset \mathbb{R}^d$ a non-empty compact set. In this case $\phi \cdot O_{\text{temp}} := \phi(O_{\text{temp}})$ and we use the topology induced by the Hausdorff distance on $\mathcal{O}$. For more in-depth information on the case of shapes seen as submanifolds of $\mathbb{R}^d$, we refer to Bauer et al.\ \cite{bauer2014overview}. Moreover, for functions $I \colon \Omega \rightarrow \mathbb{R}^n$ and $\phi \cdot I := I \circ \phi^{-1}$, $d_\mathcal{O}$ is a distance for a specific class of $\norm{\cdot}_V$ \cite{beg2005computing, miller2002metrics}. 

As mentioned above, the goal of LDDMM is to approximately solve Equations \eqref{eq:rho_I_v1} and \eqref{eq:rho_I_v2}. As the hard constraint in Equation \eqref{eq:rho_I_v2} is difficult to enforce, LDDMM introduces a soft penalty penalizing the difference between $\phi_1 \cdot O_1$ and $O_2$. In addition, this soft penalty allows us to consider objects that are not diffeomorphic to each other. The soft penalty yields the relaxed problem:
\begin{equation*}
    \begin{aligned}
        \inf_{v(\cdot, t)} \quad & \mathcal{D}(\phi_1 \cdot O_1, O_2) + \sigma^2 \int_{0}^1 \norm{v(\cdot, t)}_V^2 \mathrm{d}t  \\
        \textrm{s.t.} \quad \, & \frac{\partial \phi_t}{\partial t}(x) = v(\phi_t(x),t), \quad \phi_0(x) = x, 
    \end{aligned}
\end{equation*}
where $\mathcal{D}$ is some data fidelity term and $\sigma \in \mathbb{R}$. For instance, when dealing with images, one can use $\mathcal{D}(I_1, I_2) = \norm{I_1 - I_2}_{L_2(\Omega)}^2$ and $\phi \cdot I = I \circ \phi^{-1}$, while for meshes we can use a varifold data attachment term \cite{charon2013varifold} with $\phi \cdot \mathcal{S} := \phi(\mathcal{S})$ and $\mathcal{S}:=\{x \mid x \text{ on the shape}\}$. We note that the term $\mathcal{D}$ is very flexible. For instance, in Section \ref{sec:impl_func_choice}, we discuss a data attachment term $\mathcal{D}(O_1, O_2)$ for the case where $O_1$ is represented by an implicit representation and $O_2$ by an oriented point cloud.

\subsection{Diffeomorphic latent modeling via LDDMM PGA} \label{sec:LDDMM_PGA}
Besides registering two objects $O_1$ and $O_2$, the Riemannian distance in Theorem \ref{thm:object_dist_via_diffeomorphism_group_action} can be used for statistical latent modeling. Specifically, LDDMM is combined with principal geodesic analysis (PGA) for diffeomorphic latent modeling. In the original papers, a Bayesian approach introduces LDDMM PGA \cite{zhang2015bayesian, zhang2017probabilistic}. However, we introduce LDDMM PGA from a slightly different perspective, namely closer to the original PGA formulation in \cite{fletcher2004principal}. 

The basic idea behind PGA is to first estimate a Fréchet mean and subsequently estimate a geodesic submanifold that best explains the data. For finding the Fréchet mean within the LDDMM PGA framework, we consider the set $\mathcal{O}:=\{\phi \cdot O_{\text{temp}} \mid \phi \in G\}$ for some reference template $O_{\text{temp}}$. This reference template is arbitrary as for every $O \in \mathcal{O}$, we have $\mathcal{O} = \{\phi \cdot O \mid \phi \in G \}$, which means that any $O$ can be seen as the template. In LDDMM PGA, the Fréchet mean is a template $\mathcal{T} \in \mathcal{O}$ that minimizes the variance of the data:
\begin{equation*}
         \min_{\mathcal{T}} \expectation_{O \sim \rho}\left[d_\mathcal{O}(\mathcal{T}, O)^2\right] 
\end{equation*}
with $\rho$ the distribution of objects $O$ and $d_\mathcal{O}$ given as the distance in Equations \eqref{eq:rho_I_v1} and \eqref{eq:rho_I_v2}. Writing out $d_\mathcal{O}$ using Equation \eqref{eq:rho_I_v2} and approximating the expectation using $N$ available samples $\{O_i\}_{i=1}^N$, we get:
\begin{equation}
\begin{aligned}
    \min_{\{v_i(\cdot, t)\}_{i=1}^N, \mathcal{T}} \quad & \frac{1}{N}\sum_{i=1}^N\left[\int_0^1 \norm{v_i(\cdot, t)}_V^2 \mathrm{d} t\right] \\
     \textrm{s.t.} \quad \quad \, \, \, & \frac{\partial \phi_t^i}{\partial t}(x) = v_i(\phi_t^i(x), t), \quad \phi_0^i(x) = x, \\
    & \phi_1^i \cdot \mathcal{T}  = O_i.
   \label{eq:frechet_mean_problem_vel_vec_field_form_NEW}
\end{aligned}
\end{equation}

An optimization over a time-dependent velocity field is difficult. To circumvent this issue, we note that the geodesic optimization problems between $O_i$ and $\mathcal{T}$, which appear in Equation \eqref{eq:frechet_mean_problem_vel_vec_field_form_NEW}, are equivalent to an optimization problem over initial velocity vector fields $v_{i0}$. This equivalence is due to the diffeomorphism subgroup $G$ having a Lie group structure with Lie Algebra $V$. More precisely, if $V$ is a Hilbert space, the geodesics in $G$, which solve optimization problems \eqref{eq:rho_I_v1} and \eqref{eq:rho_I_v2}, are characterized by a Riemannian exponential map. This exponential map is characterized by the so-called EPDiff equations and maps an initial vector field $v_0 \in V$ to a time-dependent vector field $v(\cdot, t)$ that satisfies $\int_0^1 \norm{v(\cdot, t)}_V^2 \mathrm{d} t = \norm{v_0(\cdot)}_V^2$. Hence, defining $\phi_t$ as the flow of diffeomorphisms corresponding to $v(\cdot, t)$, we can construct a map $\text{exp} \colon V \rightarrow G$ as $\text{exp}{(v_0)} :=\phi_1$.  Providing an in-depth treatment of the EPDiff equation is outside the scope of this work. For more information on the EPDiff equation, we refer to Younes \cite{Younes2019, younes2007jacobi}. For our discussion here, it is sufficient to note that replacing the time-dependent velocity vector fields by the $\phi^i := \text{exp}{(v_{i,0})}$ and replacing $\int_0^1 \norm{v_i(\cdot, t)}_V^2 \mathrm{d} t$  by $\norm{v_{i, 0}}_V^2$ yields an equivalent formulation of optimization problem \eqref{eq:frechet_mean_problem_vel_vec_field_form_NEW}.

We make these adjustments and replace the hard constraint $\phi_1^i \cdot \mathcal{T}  = O_i$ by a soft penalty $\mathcal{D}(\phi^i \cdot \mathcal{T}, O_i)$:
\begin{equation}
\begin{aligned}
    \min_{\{v_{i, 0}\}_{i=1}^N, \mathcal{T}} \quad & \frac{1}{N}\sum_{i=1}^N\left[\mathcal{D}(\phi^i \cdot \mathcal{T}, O_i) + \sigma^2 \norm{v_{i, 0}}_V^2 \right] \\
    \textrm{s.t.} \quad \, \,  \, \, \, \, & \phi^i =  \text{exp}{(v_{i, 0})}, \\
\end{aligned}
\label{eq:Frechet_mean_finding_via_init_vel}
\end{equation}
where $\sigma \in \mathbb{R}$. LDDMM PGA \cite{zhang2015bayesian, zhang2017probabilistic} considers image data and optimizes the initial velocity vector fields and the template by using a grid-based discretization for $v_{i,0}$ and a finite-resolution image for the template. The exponential map on the discretized initial velocities is obtained by solving a discretized EPDiff equation. While to the best of our knowledge LDDMM PGA has not explored mesh data, Equation \eqref{eq:Frechet_mean_finding_via_init_vel} allows a mesh template and spatially discretized $v_{i,0}$ via kernel-based Reproducing Kernel Hilbert Spaces (RKHS). This approach results in an exponential map calculation that preserves the discretization \cite{charon2013varifold, ma2010bayesian, Younes2019}. 

Now that we know how to solve for the Fréchet mean in LDDMM PGA, we only need to find a geodesic submanifold that best explains the data. While we earlier mentioned that PGA \cite{fletcher2004principal} estimates the Fréchet mean first and then finds a geodesic submanifold minimizing the variance, LDDMM PGA estimates them jointly. LDDMM PGA restricts the discretized initial velocities to a linear subspace via a, possibly orthonormal, matrix $W$ and a diagonal matrix $\Lambda$. Putting this into Equation \eqref{eq:Frechet_mean_finding_via_init_vel} yields the LDDMM PGA problem:
\begin{equation}
\begin{aligned}
    \min_{W, \Lambda, \{z_i\}_{i=1}^N, \mathcal{T}} \quad & \frac{1}{N}\sum_{i=1}^N\left[\mathcal{D}(\phi^i \cdot \mathcal{T}, O_i) + \sigma^2 \norm{v_{i, 0}}_V^2 \right] \\
    \textrm{s.t.} \quad \quad \, \, \, & \phi^i =  \text{exp}{(v_{i, 0})}, v_{i, 0}=W \Lambda z_i.\\
\end{aligned}
\label{eq:LDDMM_PGA}
\end{equation}

Additional loss terms related to priors on $W$, $\Lambda$, and $z_i$ can be added to optimization problem \eqref{eq:LDDMM_PGA}. These priors relate to the standard Bayesian derivation of the LDDMM PGA model \cite{zhang2015bayesian, zhang2017probabilistic}. However, these considerations are outside the scope of this work.

\section{Riemannian Diffeomorphic Autoencoding via Implicit Neural Representations}
\label{sec:our_model}

\begin{figure}[b!]
    \centering
    \includegraphics[width=0.79\textwidth]{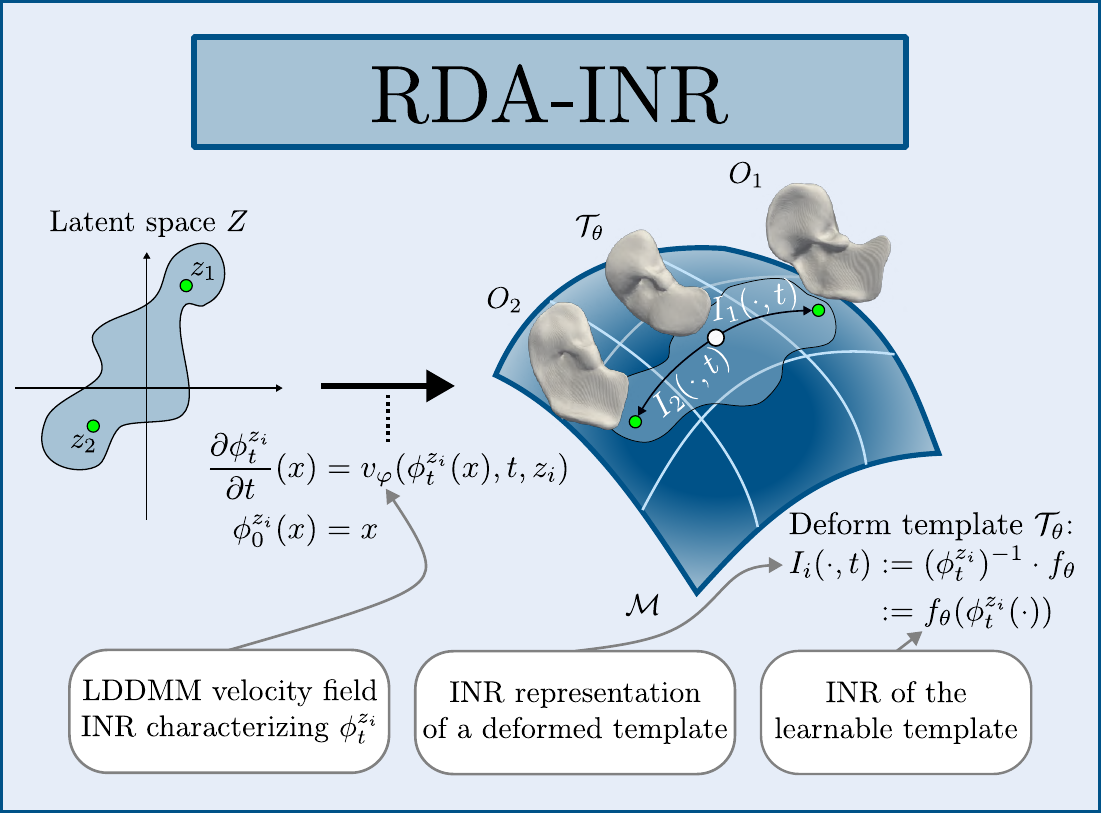} %0.79 0.61
    \caption{\textbf{RDA-INR framework} for resolution-independent and physically-consistent statistical latent modeling and atlas building. Our method maps a latent vector $z_i$ to an INR representing a time-dependent vector field $v_\varphi(\cdot, t, z_i)$. This vector field defines a flow of diffeomorphisms $\phi_t^{z_i}$ via an ODE. The diffeomorphisms $\phi_1^{z_i}$ create (reconstructed) objects $O_i$ on a subset of the Riemannian LDDMM manifold $\mathcal{M}$ by deforming a learned template $\mathcal{T}_\theta$. We obtain this deformation by parameterizing $\mathcal{T}_\theta$ with an INR $f_\theta$ and deforming $f_\theta$ via a group action: $I_i(\cdot, t) := (\phi_t^{z_i})^{-1} \cdot f_\theta$. As the template and the vector fields are parameterized by INRs, $I_i$ is a 4D INR deforming the template at $t=0$ to a reconstruction at $t=1$. The goal is to learn the template $\mathcal{T}_\theta$ as Fréchet mean of the data and to learn the $I_i(\cdot, t)$ paths as geodesics on $\mathcal{M}$ between the template $\mathcal{T}_\theta$ and the data.}
    \label{fig:method}
\end{figure}

Our model depicted in Figure \ref{fig:method} can be obtained similarly to how we derived LDDMM PGA in Section \ref{sec:LDDMM_PGA}. Specifically, we derive our model by starting with the LDDMM Fréchet mean problem \cite{joshi2004unbiased, beg2006computing, ma2008bayesian, ma2010bayesian}. 

As shown in Section \ref{sec:LDDMM_PGA}, if we want to find the Fréchet mean based on LDDMM, we should solve the problem in Equation \eqref{eq:frechet_mean_problem_vel_vec_field_form_NEW}. In LDDMM PGA, one first parameterizes the time-dependent diffeomorphisms via an initial velocity field, replaces the hard constraint with a soft penalty, and subsequently discretizes the template and the initial velocity vector fields. For our model, we skip the initial velocity field parameterization and immediately replace the hard constraint with a soft constraint using a general data-fitting term $\mathcal{D}$ and a $\sigma \in \mathbb{R}$:
\begin{equation}
\begin{aligned}
    \min_{\{v_i(\cdot, t)\}_{i=1}^N, \mathcal{T}} \quad & \frac{1}{N}\sum_{i=1}^N\left[\mathcal{D}\left(\phi_1^i \cdot \mathcal{T}, O_i\right) + \sigma^2\int_0^1 \norm{v_i(\cdot, t)}_V^2 \mathrm{d} t\right] \\
    \textrm{s.t.} \quad \quad \, \, & \frac{\partial \phi_t^i}{\partial t}(x) = v_i(\phi_t^i(x), t), \quad \phi_0^i(x) = x. 
   \label{eq:model_derivation_eq1}
\end{aligned}
\end{equation} 
Here different data fidelity terms $\mathcal{D}$ can be chosen depending on the data.  

For LDDMM PGA, we required a grid-based or spatial discretization of the template and initial velocity vector fields to solve the optimization problem in Equation \eqref{eq:model_derivation_eq1}. Furthermore, the chosen discretization allows the restriction of the initial velocity vector fields to a linear subspace. Our approach circumvents the need for grid-based or spatial discretizations by parameterizing the template and time-dependent velocity vector fields via INRs. Specifically, we parameterize $\phi_1^i$ by specifying the corresponding flow in reverse time. Defining $z_i \in \mathbb{R}^{d_z}$ as the latent code belonging to the object $O_i$ and $v_\varphi\colon\Omega \times [0,1] \times \mathbb{R}^{d_z} \rightarrow \mathbb{R}^d$ as an implicit neural representation, the reverse time flow is given by $\frac{\partial }{\partial t}\phi_t^{z_i}(x) = v_\varphi(\phi_t^{z_i}(x), t, z_i)$. Jointly learning the weights $\varphi$ and the latent codes $z_i$ gives:
\begin{equation}
\begin{aligned}
    \min_{\varphi, \{z_i\}_{i=1}^N, \mathcal{T}} \quad & \frac{1}{N}\sum_{i=1}^N\left[\mathcal{D}((\phi_1^{z_i})^{-1} \cdot \mathcal{T}, O_i)  + \sigma^2\int_0^1 \norm{v_\varphi(\cdot, t, z_i)}_V^2 \mathrm{d} t\right] \\
    \textrm{s.t. } \quad \, \, \, \, &  \frac{\partial \phi_t^{z_i}}{\partial t}(x) = v_\varphi(\phi_t^{z_i}(x), t, z_i), \quad \phi_0^{z_i}(x) = x.
\end{aligned}
\label{eq:nn_optimization_problem_without_INR_template}
\end{equation}

\noindent Finally, we represent the template $\mathcal{T}$ by an implicit neural representation $f_\theta\colon\Omega \rightarrow \mathbb{R}$ to obtain a resolution-independent method. The group action on the implicit neural representation is:
\begin{equation}
    I(x, z, t) := ((\phi_t^z)^{-1} \cdot f_\theta)(x) := f_\theta(\phi_t^z(x)),
    \label{eq:impl_func_decoder}
\end{equation}
where $I(x, z, 0)$ equals the implicit representation of the template and $I(x, z, 1)$ is the reconstructed implicit representation of an object $O$. This group action motivates parameterizing the flow in reverse time as we need to evaluate $f_\theta$ on the reverse-time flow. Alternatively, the group action can be associated with a group action on the set-based definition of the template shape $\mathcal{T}_\theta := \{x \mid f_\theta(x)=0\}$, namely $\phi \cdot \mathcal{T}_\theta := \phi(\mathcal{T}_\theta)$. Both group actions yield a strategy to obtain a reconstructed mesh or point cloud for the reconstructed shape: apply marching cubes on $I(x, z, 1)$ or use $(\phi_{1}^z)^{-1} \cdot \mathcal{T}_\theta = (\phi_{1}^z)^{-1}(\mathcal{T}_\theta)$. Incorporating the template $\mathcal{T}_\theta$ and the associated group action into Equation \eqref{eq:nn_optimization_problem_without_INR_template} and assuming some $\mathcal{D}$ for fitting the deformed template or its implicit representation to the data, yields the final optimization problem:
\begin{equation}
\begin{aligned}
    \min_{\theta, \varphi, \{z_i\}_{i=1}^N} \quad & \frac{1}{N}\sum_{i=1}^N\left[\mathcal{D}((\phi_1^{z_i})^{-1} \cdot \mathcal{T}_\theta, O_i)  + \sigma^2\int_0^1 \norm{v_\varphi(\cdot, t, z_i)}_V^2 \mathrm{d} t\right] \\
    \textrm{s.t. } \quad \, \, \,&  \frac{\partial \phi_t^{z_i}}{\partial t}(x) = v_\varphi(\phi_t^{z_i}(x), t, z_i), \quad \phi_0^{z_i}(x) = x.
\end{aligned}
\label{eq:NN_optimization_problem}
\end{equation}

Comparing Equation \eqref{eq:NN_optimization_problem} with the LDDMM PGA optimization problem in Equation \eqref{eq:LDDMM_PGA}, we see several similarities and differences. In both problems, we optimize over latent vectors $z_i$ and weights to define diffeomorphisms. However, in LDDMM PGA, the weights construct a linear relationship between the latent vectors and discrete initial velocity vector fields, while the weights in our model parameterize neural networks. This parameterization yields a nonlinear relationship between latent vectors and the time-dependent velocity vector fields. Compared to the linear relationship, the nonlinear relationship makes the latent space more difficult to navigate as it is a-priori unclear how a change in latent space affects the velocity vector fields. Moreover, our model does not use a grid-based or spatial discretization for the velocity vector fields and the template. Instead, our model parameterizes them via INRs, making the model resolution independent. By optimizing the weights of the INRs, the model searches over finite-dimensional nonlinear submanifolds of full infinite-dimensional spaces. Specifically, within these submanifolds, our model searches for the best template and nonlinear submanifold of time-dependent velocity vector fields. Finally, due to the nonlinear relationship between latent vectors and velocity vector fields, and the INR representation of the vector fields, it is not possible to enforce the orthogonality properties of LDDMM PGA using an orthogonal weight matrix $W$ as shown in Equation \eqref{eq:LDDMM_PGA}.

Up until now, we kept the data fidelity term $\mathcal{D}$ general such that the optimization problem in Equation \eqref{eq:NN_optimization_problem} is as close as possible to the general LDDMM PGA model. For the point cloud or mesh data that we consider, it is not immediately obvious what data fidelity $\mathcal{D}$ to use in combination with our INR representation of the template. We introduce the used data fidelity term for this data in Section \ref{sec:impl_func_choice}. Subsequently, Section \ref{sec:reg_term} introduces the velocity field parameterization and the chosen $\norm{\cdot}_V$. Finally, we treat how we obtain latent codes for unseen instances in Section \ref{sec:recon_autodec}. For details about the neural network architectures of $f_\theta$ and $v_\varphi$ and how we numerically deform the template to obtain a reconstruction, see Appendix \ref{app:impl_details}.

\subsection{Data fidelity term for shape data} \label{sec:impl_func_choice}
When dealing with point cloud or mesh data, recent resolution-independent methods employ a deformable template for joint shape encoding and registration \cite{zheng2021deep, sun2022topology} and use $\mathcal{D}(I_1, I_2) = \norm{I_1 - I_2}_{L_1(\Omega)}$ with $I_i$ a ground truth signed distance representation of the point cloud or mesh. However, there are three issues with such a $\mathcal{D}$: (i) signed distance functions of diffeomorphic shapes are not always diffeomorphic, (ii) if they are diffeomorphic, they influence the diffeomorphism that we find, and (iii) signed distance functions are not always available for point clouds or are non-trivial to obtain.

\begin{figure}[t!]
    \centering
    \includegraphics[width=0.55\textwidth]{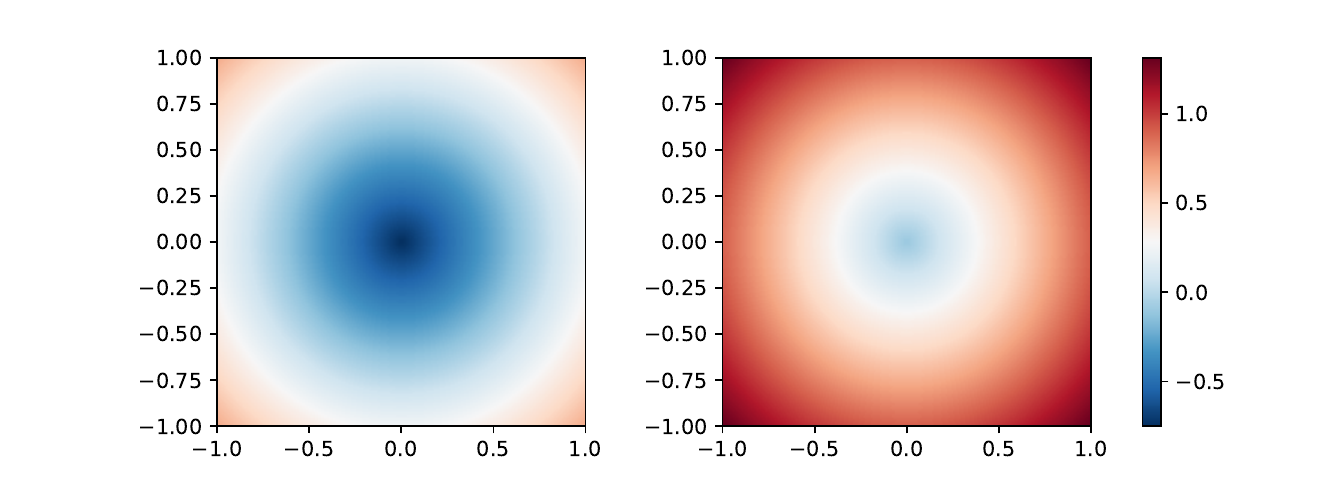}
    \caption{SDF values in $[-1,1]^2$ of a circle with radius $0.75$ (left) and of a circle with radius $0.1$ (right).}
    \label{fig:sdf_heatmap_circles}
\end{figure}

To illustrate the first issue, take two circles of radii $0.75$ and $0.1$, respectively. Moreover, represent them by their SDF, as presented in Equation \eqref{eq:SDF}. These SDFs are shown in Figure \ref{fig:sdf_heatmap_circles}. We have $\min_{x \in \Omega} \text{SDF}_{\mathcal{S}_0}(x) < \min_{x \in \Omega} \text{SDF}_{\mathcal{S}_1}(x)$ as dark blue is present in the figure of the circle with radius $0.75$ but not in the other figure. Consequently, no diffeomorphism matching the two implicit representations (i.e., SDF images) exists, which contradicts Equation \eqref{eq:NN_optimization_problem} where both SDFs need to be diffeomorphically matched to the same template implicit function.

For the second issue, assume shapes $\mathcal{S}_0$ and $\mathcal{S}_1$ are represented by implicit functions that can be exactly matched. If a point $x_0$ in shape $\mathcal{S}_0$ has implicit function value $s$, it must be matched with a point $x_1$ in shape $\mathcal{S}_1$ that also has implicit function value $s$. By also matching points outside and inside the shape based on information like the SDF value, we influence how the points on the shape are matched.

To overcome these issues, we use the approach used by Sitzmann et al.\ \cite{sitzmann2020implicit} and Gropp et al.\ \cite{gropp2020implicit} for learning an SDF from point cloud and mesh data. This approach uses the point cloud representation for training shapes $\{\mathcal{S}_i\}_{i=1}^N$ and uses the following data fidelity term $\mathcal{D}$:
\begin{equation}
    \begin{aligned}
    \mathcal{D}((\phi_1^z)^{-1} \cdot \mathcal{T}_\theta, \mathcal{S}_i) := \expectation_{x\in \mathcal{S}_i} \left[ \lvert I(x, z, 1) \rvert  \right. & \left. + \tau \left(1 -F_{\text{cos}}(\nabla_x I(x,z,1), n_i(x)) \right) \right] \\
    & + \beta \expectation_{x \in \Omega \setminus \mathcal{S}_i}\left[\exp{(-\alpha \lvert I(x,z_i,1) \rvert)}\right],
    \end{aligned}
    \label{eq:data_fitting_term_pointclouds}
\end{equation}
with $I(x, z, t)$ given in Equation \eqref{eq:impl_func_decoder}, $\tau, \beta, \alpha \in \mathbb{R}$, $n_i(x)$ the unit normal of the shape $\mathcal{S}_i$ at $x\in \mathcal{S}_i$, and $F_{\text{cos}}$ the cosine similarity:
\begin{equation*}
    F_{\text{cos}}(a,b) := \frac{\langle a, b \rangle_2}{\max(\norm{a}_2 \cdot \norm{b}_2, 10^{-8})},\quad a,b\in\mathbb{R}^d.
\end{equation*}
This data-fitting term solves the previously mentioned issues since the first expectation only focuses on matching points on the training shapes with points on the template shape. In other words, all points on the training shapes are matched to the zero level set of the INR $f_\theta$. However, using only the first expectation encourages $f_\theta = 0$ as this ensures all points are matched to the zero level set of $f_\theta$. To avoid this trivial template, the loss term involving the $\beta$ parameter ensures that only points on $\mathcal{S}_i$ are matched to the zero level set of $f_\theta$. In addition, to enforce additional structure on the learned template INR and to further ensure $f_\theta \neq 0$, we add an eikonal penalty \cite{sitzmann2020implicit, deng2021deformed, gropp2020implicit} with regularization constant $\lambda \in \mathbb{R}$ to optimization problem \eqref{eq:NN_optimization_problem}, which regularizes the template INR to be a signed distance function as in Equation \eqref{eq:SDF}. Finally, we add an isotropic Gaussian prior with regularization constant $\gamma$ on the latent codes. Combining all loss functions results in the final optimization problem:
\begin{equation}
\begin{aligned}
    \min_{\theta, \varphi, \{z_i\}_{i=1}^N} \quad &  \frac{1}{N}\sum_{i=1}^N  
    \left[ \expectation_{x\in \mathcal{S}_i}\left[ \lvert I(x,z_i,1) \rvert + \tau \left(1- F_{\text{cos}}(\nabla_x I(x,z_i,1), n_i(x)) \right)  \right] \right. 
      \\
     & \left. \qquad \quad + \beta \expectation_{x \in \Omega \setminus \mathcal{S}_i}\left[\exp{(-\alpha \lvert I(x,z_i,1) \rvert)}\right] + \sigma^2 \int_0^1 \norm{v_\varphi(\cdot, t, z_i)}_V^2 \mathrm{d} t + \gamma \norm{z_i}_2^2\right] \\
     & + \lambda \expectation_{x\in \Omega}\left [ \lvert \norm{\nabla_x f_\theta(x)}_2 - 1 \rvert\right] \\
    \textrm{s.t. } \quad \, \, \, & \frac{\partial \phi_t^{z_i}}{\partial t}(x) = v_\varphi(\phi_t^{z_i}(x), t, z_i), \quad \phi_0^{z_i}(x) = x, \\
    & I(x, z, t) := ((\phi_t^z)^{-1} \cdot f_\theta)(x) := f_\theta(\phi_t^z(x)).
\end{aligned}
\label{eq:NN_optimization_problem_IGR}
\end{equation}

\begin{remark}
Another approach to solve the issues is using the occupancy function in Equation \eqref{eq:OccFunc} as the implicit representation for the data and the template shape. Earlier works using occupancy functions as implicit representations are Mescheder et al.\ \cite{mescheder2019occupancy} and Niemeyer et al.\ \cite{niemeyer2019occupancy}. The reason occupancy functions solve the issues is that we only match points inside (outside) shape $\mathcal{S}_0$ to points inside (outside) shape $\mathcal{S}_1$ and do not take into account information like the SDF value. As the occupancy function and the occupancy values resemble probabilities, we can use the binary cross entropy loss as data fidelity term $\mathcal{D}$ in optimization problem \eqref{eq:NN_optimization_problem} \cite{mescheder2019occupancy, niemeyer2019occupancy}. However, as shown in Appendix \ref{app:occnet_vs_IGR}, our strategy that uses point clouds as data representation allows for higher-quality reconstructions. This result further motivates the use of the point cloud representation.
\end{remark}
\begin{remark}
    Although we specialize our model to point cloud and mesh data representing shapes, our model can easily be extended to images that do not necessarily represent shapes. Specifically, in Equation \eqref{eq:NN_optimization_problem}, we can choose $\mathcal{D}(I_1, I_2)$ as any data fitting attachment for images and use $\mathcal{T}_\theta = f_\theta$ with $\phi \cdot \mathcal{T}_\theta = \mathcal{T}_\theta \circ \phi^{-1}$.
\end{remark}

\subsection{Choice of velocity field parameterization and velocity field regularization term} \label{sec:reg_term}
As parameterization of $v_\varphi$, we choose a quasi-time varying velocity vector field defined by several INRs representing stationary velocity vector fields. For more details, we refer to Appendix \ref{app:impl_details}. For the norm $\norm{\cdot}_V$ in optimization problems \eqref{eq:NN_optimization_problem} and \eqref{eq:NN_optimization_problem_IGR} we choose an isometric (rigid) deformation prior on the velocity vector fields by combining the Killing energy \cite{solomon2011killing, tao2016near} with an $L_2(\Omega)$ penalty:
\begin{equation}
    \norm{\nu}_V^2 := \int_\Omega \lVert (J \nu) + (J \nu) ^T \rVert_F^2 + \eta \norm{\nu}^2_2\mathrm{d} x,
    \label{eq:Killing_energy}
\end{equation}
where $\nu$ is a velocity vector field, $J \nu$ the Jacobian of $\nu$ with respect to $x$, and $\eta \in \mathbb{R}$. If $\eta$ and the functional in Equation \eqref{eq:Killing_energy} are small, then we expect the template to deform almost isometrically to the reconstructed objects via $I(x, z, t)$ in Equation \eqref{eq:impl_func_decoder}. In the numerical experiments, we use this property and a dataset with a rigidity prior to highlight the importance of a data-consistent regularizer. Furthermore, this property is interesting as it allows us to jointly perform affine registration and LDDMM registration. The regularizer and its property are novel in the LDDMM literature. In Appendix \ref{SM:kill_ener_as} we showcase that our regularizer provides a novel motivation and special case of a lesser-used norm in the LDDMM literature. We also show that our regularizer adds a rigidity prior to a more standard Sobolev LDDMM norm.

\subsection{Encoding objects} 
\label{sec:recon_autodec}
We use the strategy from DeepSDF \cite{park2019deepsdf} to encode (new) objects. More precisely, we solve the following optimization problem for encoding the object $\mathcal{S}$:
\begin{equation}\label{eq:autodecoder_reconstruction}
    \min_z  \mathcal{D}_{rec}((\phi_1^z)^{-1} \cdot \mathcal{T}_\theta, \mathcal{S}) + \gamma \norm{z}_2^2,
\end{equation}
where $\gamma \in \mathbb{R}$ corresponds to an isotropic Gaussian prior on the latent codes, $\phi_t^z$ is defined via the ODE in optimization problems \eqref{eq:NN_optimization_problem} and \eqref{eq:NN_optimization_problem_IGR}, $\mathcal{T}_\theta$ is the learned template, and $\mathcal{D}_{rec}$ is $\expectation_{x\in \mathcal{S}} \left[ \lvert I(x, z, 1) \rvert \right]$ with $I(x,z,t)$ given by Equation \eqref{eq:impl_func_decoder} and $\mathcal{S}$ is represented by a mesh or a point cloud. Note that this $\mathcal{D}_{rec}$ differs from  $\mathcal{D}$ in Equation \eqref{eq:data_fitting_term_pointclouds}. As the latent space generates a prior on the data, reconstructing solely with the point cloud vertices is sufficient.

\begin{remark}
Our encoding procedure presented in Equation \eqref{eq:autodecoder_reconstruction} is similar to the encoding strategy in PGA \cite{fletcher2004principal}. In PGA, the first step is to find the Fréchet mean $\mu \in M$ of the data lying on some manifold $M$. Subsequently, PGA identifies a subspace $W$ of the tangent space $T_\mu M$ such that most variability in the data is described by $H = \text{exp}_\mu(W)$, where $\text{exp}_\mu$ is the exponential map at $\mu$. Finally, to project a data point $p\in M$ onto $H$, one calculates:
\begin{equation*}
    \argmin_{w\in W} d(p,\text{exp}_\mu(w))^2.
\end{equation*}

Equation \eqref{eq:autodecoder_reconstruction} resembles this optimization problem. Similarly to the relationship between Equation \eqref{eq:NN_optimization_problem} and LDDMM PGA, $(\phi_1^z)^{-1} \cdot \mathcal{T}_\theta$ should approximate the exponential map and instead of searching over a linear vector space $W$, we search over a nonlinear submanifold of vector fields defined via a latent space. Finally, although $\mathcal{D}_{rec}$ is not a Riemannian distance, it replaces the Riemannian distance $d$, as also done in, e.g., Charlier et al.\ \cite{charlier2018distortion}.
\end{remark}

\section{Numerical results} \label{sec:numerical_results}
In this section, we demonstrate the benefit of the physical consistency and resolution independence of our model. To evaluate the benefit of physical consistency, we compare the Riemannian LDDMM regularizer to a non-Riemannian regularizer. We utilize two shape datasets for this comparison: a synthetic rectangles dataset and a liver dataset \cite{sun2022topology, chen2021deep}. Subsequently, we compare our method to the diffeomorphic autoencoder (DAE) \cite{bone2019learning} on the liver dataset to illustrate the benefit of resolution independence for LDDMM statistical latent modeling. For details about the data source, the data preprocessing, and the training, we refer to Appendix \ref{app:impl_details}.

\subsection{Physical consistency and statistical shape modeling via latent spaces}
To assess whether physical consistency is desirable for statistical shape modeling via latent spaces, we compare our model that regularizes the flow via the Riemannian LDDMM regularization (see Equation \eqref{eq:NN_optimization_problem_IGR}) to our model utilizing the non-Riemannian pointwise regularization, which is presented in Appendix \ref{app:pointwise_loss}. The latter model is the state-of-the-art baseline model by Sun et al.\ \cite{sun2022topology} with a different data-fitting term and a different approach to solving the ODE. In Appendix \ref{app:comparison_RDA-INR_NDF}, we show that our changes do not deteriorate the reconstruction performance. Furthermore, in Appendix \ref{app:pointwise_loss}, we explain the relationship of the non-Riemannian regularization with the Riemannian LDDMM regularization and explain why we call it a non-Riemannian regularization. Essentially, the pointwise regularization, like our Riemannian regularization, penalizes deviations of the time-dependent diffeomorphism at the final time from the identity diffeomorphism. However, the intermediate time diffeomorphisms are not promoted to correspond to a Riemannian geodesic between the identity and final diffeomorphism. 

We assess the mean-variance analysis capabilities of the models by visualizing the learned templates and deformations from the template to the training shapes. We relate our findings to Figure \ref{fig:possible_problem_with_mean_and_variance_calculation}. Subsequently, we discuss reconstruction generalization and robustness of the reconstruction procedure to noisy data. This discussion shows the effect of Riemannian regularization on the quality and stability of the reconstruction procedure. 

\subsubsection{Learning shape Fréchet means} \label{sec:template_exps}
\begin{figure}[b!]
    \centering
    
    \subfloat[Liver Riemannian regularization.]{ \label{fig:templates_subfigure_a}\includegraphics[width=0.23\textwidth]{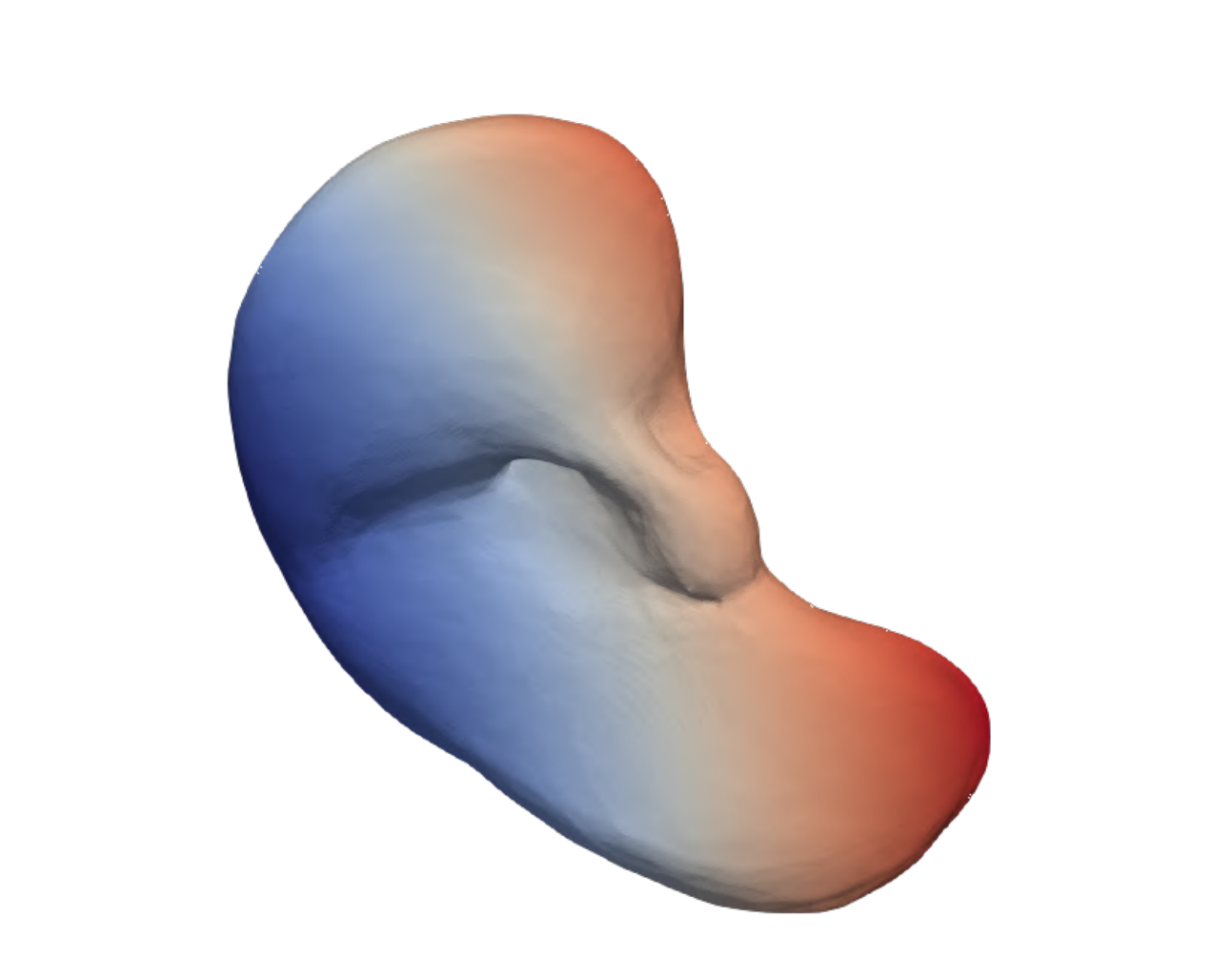}}
    \hfill
    \subfloat[Liver non-Riemannian regularization.]{ \label{fig:templates_subfigure_b}\includegraphics[width=0.23\textwidth]{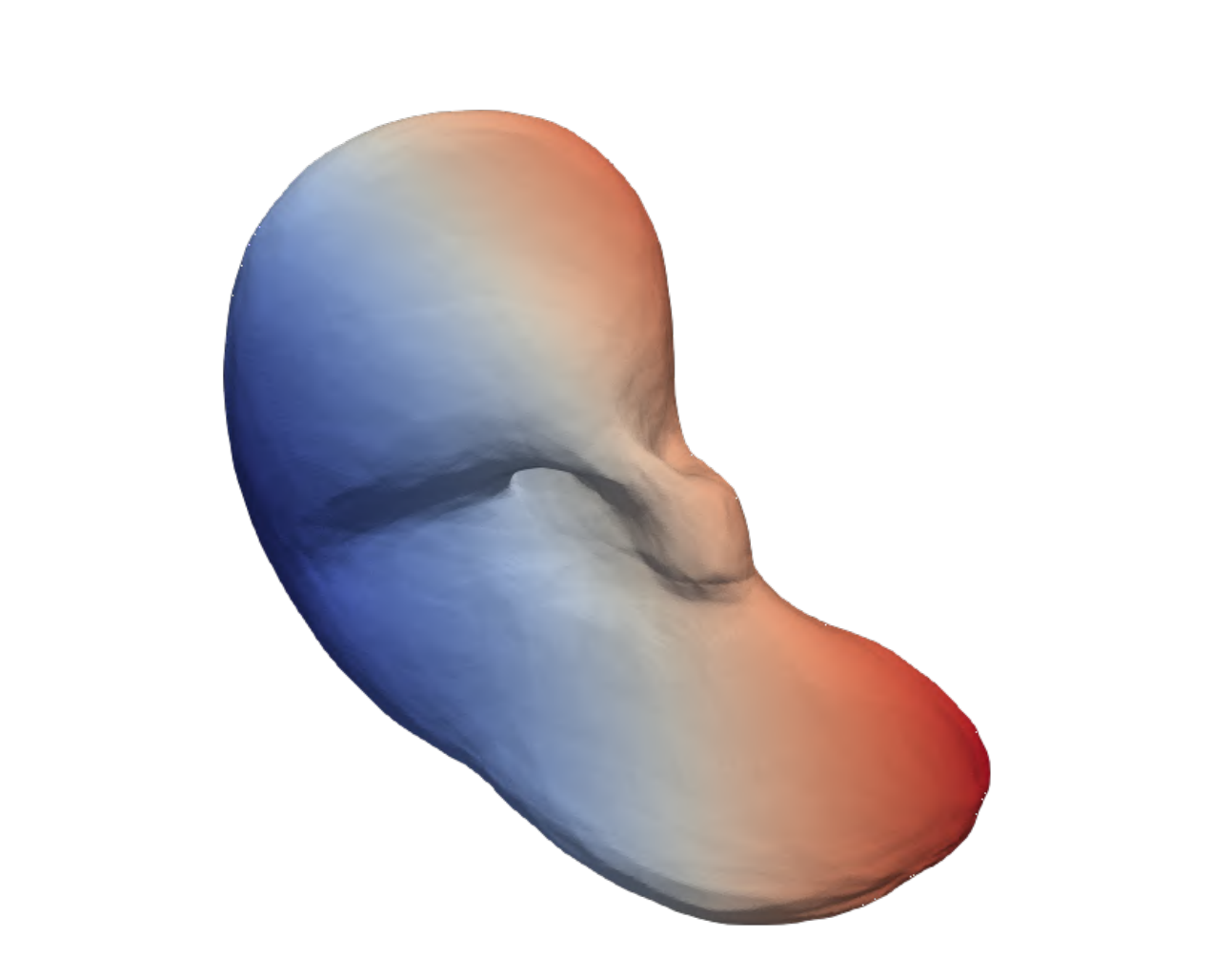}}
    \hfill
    \subfloat[Rect. Riemannian regularization.]{ \label{fig:templates_subfigure_c}\includegraphics[width=0.23\textwidth]{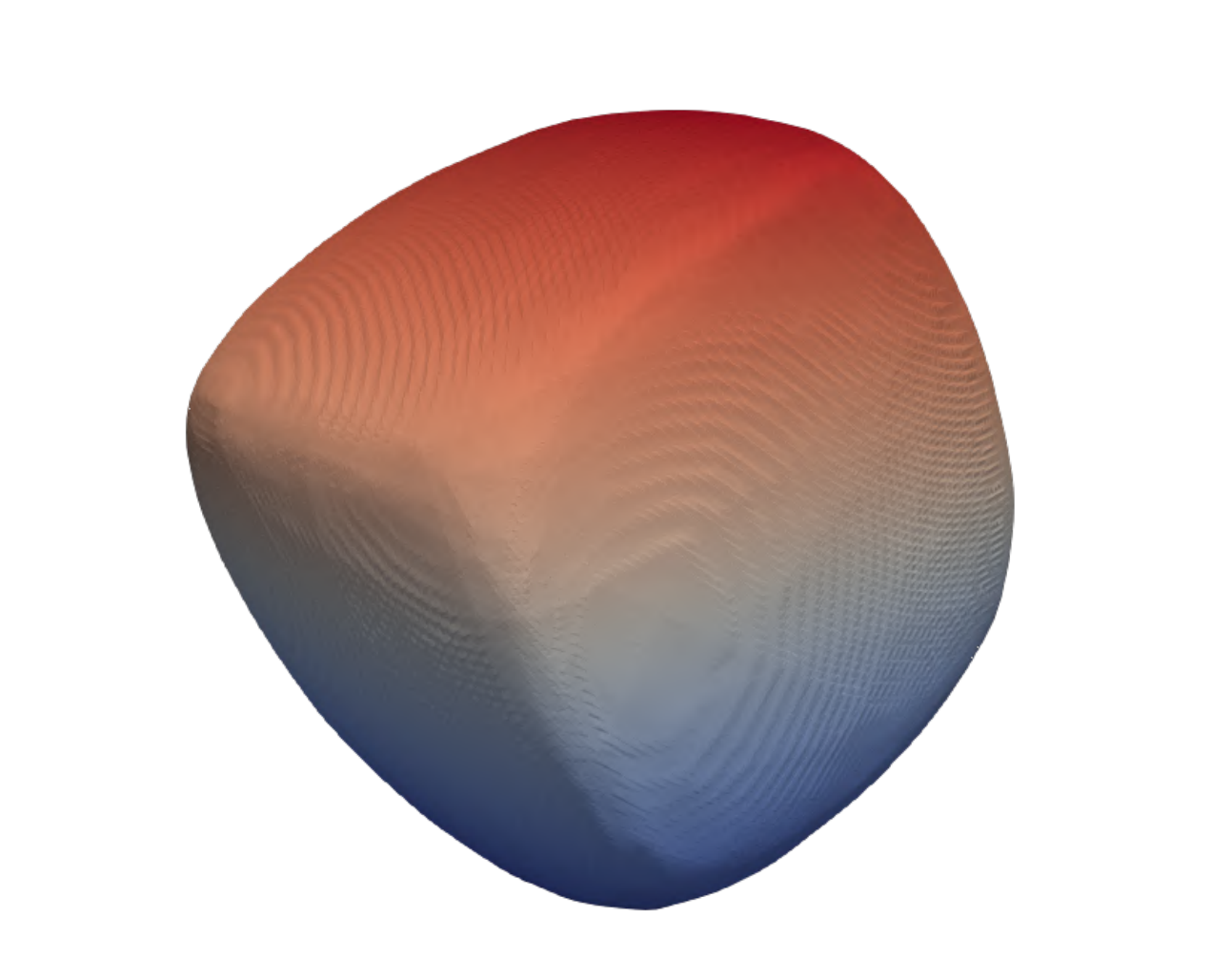}}
    \hfill
    \subfloat[Rect. non-Riemannian regularization.]{ \label{fig:templates_subfigure_d}\includegraphics[width=0.23\textwidth]{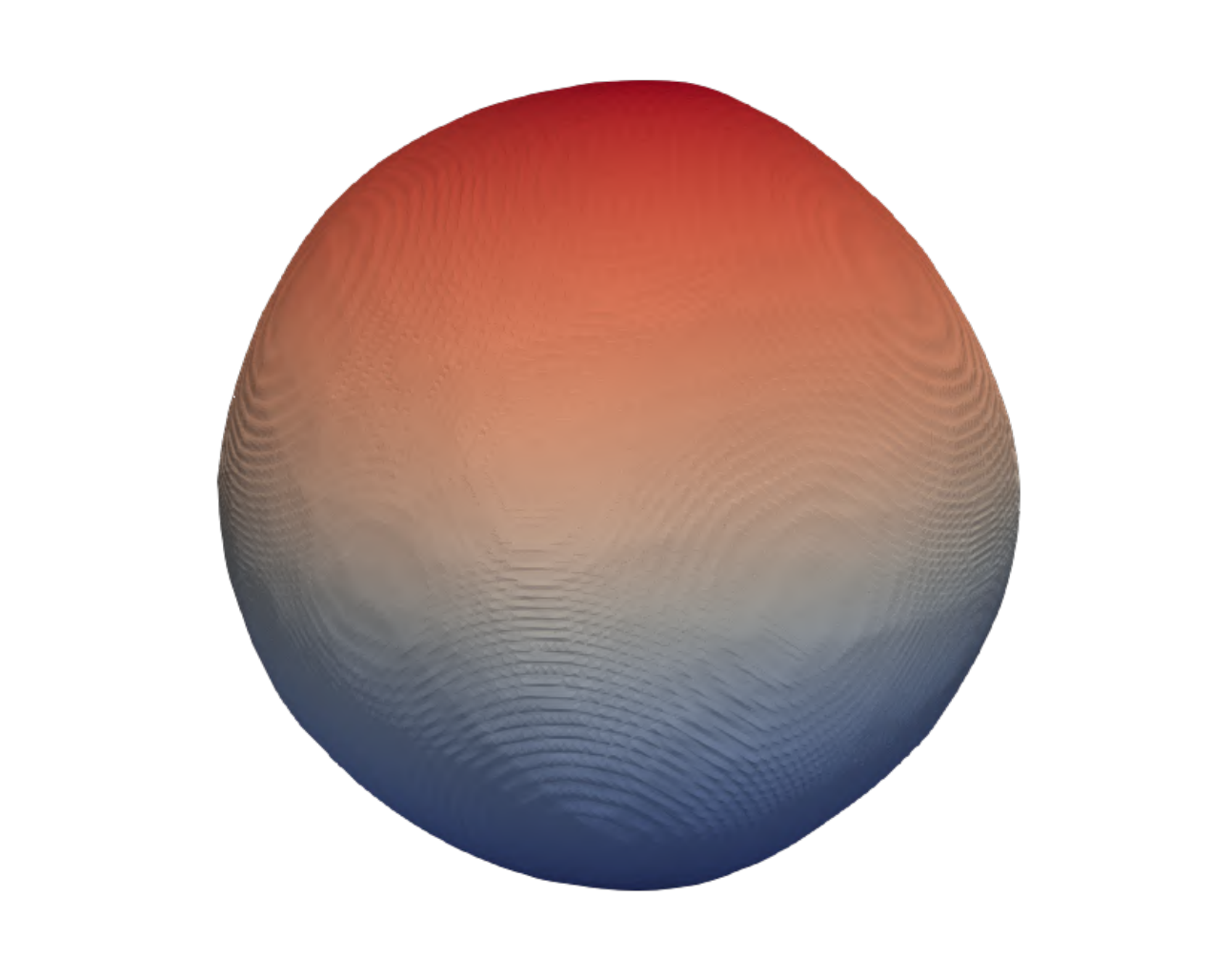}}
        \caption{\textbf{Template (atlas) building.} The learned templates of two different models trained on the rectangles (Rect.) dataset and the liver dataset. We use the model learned using the Riemannian LDDMM regularization (a and c)  and the model learned using the non-Riemannian pointwise loss  (b and d). We use $\eta = 0.05$ and $\eta=50$ in Equation \eqref{eq:Killing_energy} for the rectangles and liver dataset, respectively.}
        \label{fig:templates}
\end{figure}

For training the models with Riemannian LDDMM regularization, we need to pick a value for $\eta$. We choose $\eta = 0.05$ for the rectangles dataset as we expect rigid body motions from the template to the training shapes. For the liver data, the only prior that we have is that the liver should not change too quickly. Hence, we pick $\eta=50$ such that $\norm{\cdot}_V \approx \eta \norm{\cdot}_{L_2(\Omega)}$.

After training, we obtain the template shapes in Figure \ref{fig:templates}. While the templates of the liver dataset look almost identical, the templates for the rectangles dataset differ. In particular, both liver templates resemble livers, whereas the template learned on the rectangles data only resembles the data for the Riemannian model. The spherical template in Figure \ref{fig:templates_subfigure_d} is not desired as it lies outside the data distribution, as depicted in the third category in Figure \ref{fig:possible_problem_with_mean_and_variance_calculation}.

The difference between the liver and rectangle datasets stems from the difference in the prior. For the liver dataset, there is no clear deformation prior besides that the liver should not change too quickly. The non-Riemannian regularization induces this prior by construction while the Riemannian regularization enforces this by our choice of $\eta = 50$. More precisely, as $\norm{\cdot}_V \approx \eta \norm{\cdot}_{L_2(\Omega)}$ and as the pointwise loss can be interpreted as a non-Riemannian version of the LDDMM regularization with $\norm{\cdot}_V = \norm{\cdot}_{L_2(\Omega)}$ (see Appendix \ref{app:pointwise_loss}), both regularizations induce the same prior on the deformation, yielding similar templates in Figures \ref{fig:templates_subfigure_a} and \ref{fig:templates_subfigure_b}.

While the liver shapes should only change gradually, the rectangles should also move rigidly. Consequently, we let the Killing energy play a more prominent role in $\norm{\cdot}_V$ in the Riemannian LDDMM regularization. This choice yields almost rigid deformations from the template to the training shapes, which results in the square template in Figure \ref{fig:templates_subfigure_c}. As the pointwise loss does not induce an isometry prior and is a non-Riemannian version of the LDDMM regularization with $\norm{\cdot}_V = \norm{\cdot}_{L_2(\Omega)}$, we obtain the spherical template in Figure \ref{fig:templates_subfigure_d}.

In summary, while on the liver dataset, the templates look similar, the Riemannian regularization is required for the rectangles dataset to learn a template that fits the data distribution. Hence, the Riemannian LDDMM regularization allows for more flexibility in learning a template shape that resembles the data. Moreover, on data with a clear Riemannian prior that not only promotes small $\norm{v}_{L^2(\Omega)}$, such as a rigidity prior, the Riemannian LDDMM regularization is preferred.

\subsubsection{Learning geodesic shape deformations} \label{sec:template_deformation}

\begin{figure}[b!]
    \centering 
    \includegraphics[width=0.76\textwidth]{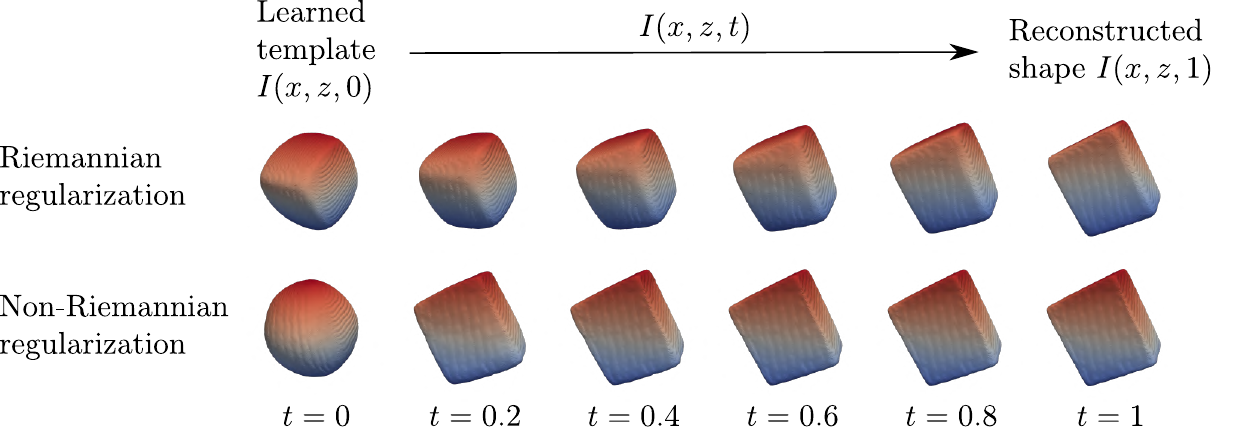}
    \caption{\textbf{Template deformations}. The deformation of the template into a reconstructed rectangle using the model learned with the Riemannian LDDMM regularization (top) and with the model learned using the non-Riemannian pointwise loss (bottom). The colors represent the matching of the points to their template.}
    \label{fig:deformation_flow_rectangles}
\end{figure}

\begin{figure}[b!]
    \centering 
    \includegraphics[width=0.76\textwidth]{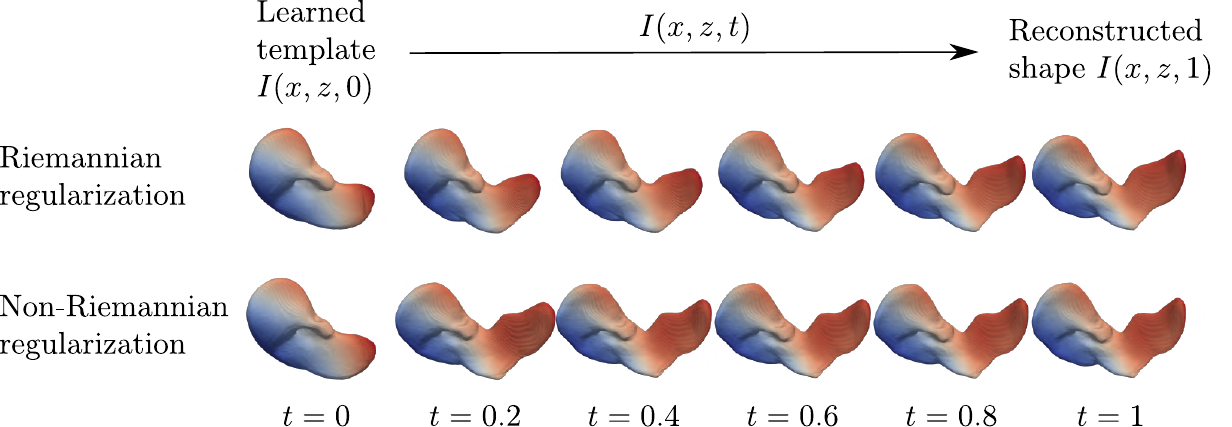}
    \caption{\textbf{Template deformations.} The deformation of the template into a reconstructed liver using the model learned with the Riemannian LDDMM regularization (top) and with the model learned using the non-Riemannian pointwise loss (bottom). The colors represent the matching of the points to their template.}
    \label{fig:deformation_flow_liver}
\end{figure}

The previous section shows a regularizer's effect on the learned Fréchet mean. To finalize the discussion on Figure \ref{fig:possible_problem_with_mean_and_variance_calculation}, we assess the variance calculation by qualitatively assessing whether the template deformations are geodesics.  

In Figure \ref{fig:deformation_flow_rectangles} the template is transformed into a reconstructed training shape for each model trained on the rectangles dataset. We notice that the non-Riemannian model quickly transitions from the spherical template to the target shape and then has a tendency to stay in a nearly identical configuration. In contrast, the model with the Riemannian LDDMM regularization smoothly rotates and scales the rectangular template shape to the final reconstructed shape. Hence, it fits the rigid body motion prior induced by the Killing energy. 

\begin{figure}[t!]
    \centering 
    \includegraphics[width=0.76\textwidth]{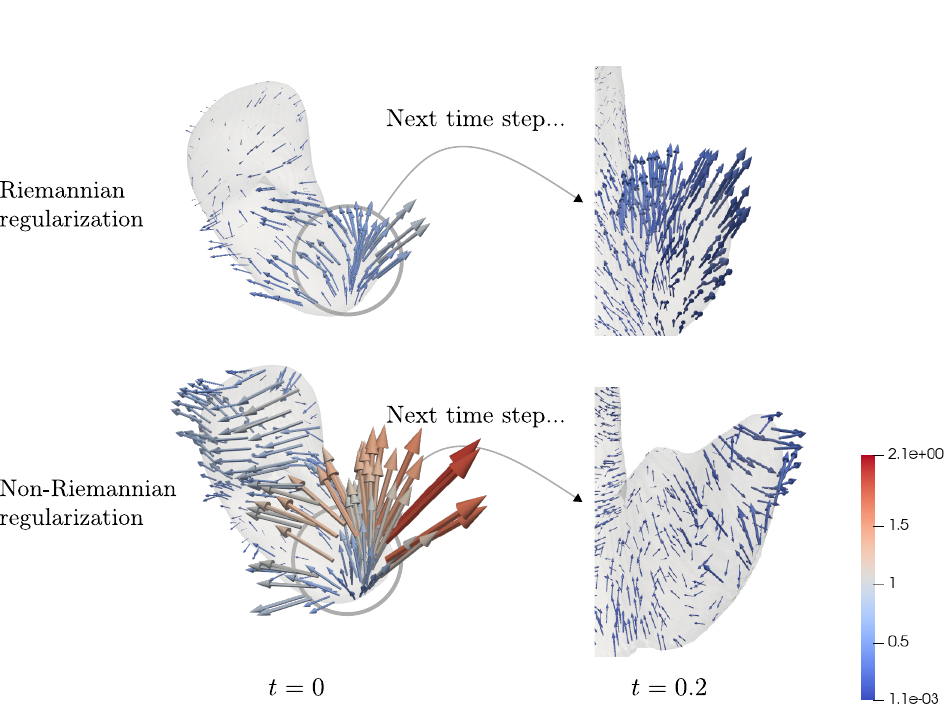}
    \caption{\textbf{Velocity vector fields.} The velocity vector fields at $t=0$ and $t=0.2$ (zoom-in) of the deformations in Figure \ref{fig:deformation_flow_liver}. The colors represent the magnitude of the vector field. The non-Riemannian model produces an unrealistic magnitude at $t=0$ and an inconsistent vector field at $t=0.2$ as a consequence. }
    \label{fig:deformation_flow_liver_vel_field}
\end{figure}

Regarding the liver dataset, Figure \ref{fig:deformation_flow_liver} shows that the deformation with the Riemannian LDDMM regularization is again smoother in time than the deformation with the non-Riemannian pointwise loss. Figure \ref{fig:deformation_flow_liver_vel_field} reinforces this as it shows that the vector field of the non-Riemannian model at time $t=0.0$ is everywhere much larger than the vector field of the Riemannian model. Moreover, at $t=0.2$, the vector field of the Riemannian model points in approximately the same direction as at $t=0.0$, while this does not apply to the non-Riemannian model. Hence, the vector field of the Riemannian model is smoother in time.

To explain the preceding observations, we note that, by construction, the non-Riemannian pointwise loss encourages a rapid transition to the target shape followed by a tendency to stay in a nearly identical configuration. Consequently, the velocity vector fields at later times $t$ fine-tune the obtained reconstructions. Hence, the velocity vector field at $t=0.2$ does not necessarily point in approximately the same direction as the velocity vector field at $t=0.0$, as shown in Figure \ref{fig:deformation_flow_liver_vel_field}. In contrast, the Riemannian LDDMM regularization penalizes rapid transitions and prefers to gradually get closer to the target shape in a prior-consistent manner. Hence, the Riemannian regularization allows for a smooth, physically plausible template deformation into the reconstructed shapes.

In Table \ref{tab:summary_template_and_deformation_experiment}, we summarize the discussion regarding Fréchet means and variance. The table shows that the Riemannian LDDMM regularization is more flexible in influencing the template than the non-Riemannian pointwise loss. Moreover, to obtain good variance estimates, the LDDMM regularization is needed to obtain geodesic deformations of the template to the reconstructions. Hence, the model with the Riemannian LDDMM regularization is the only model in category 1 of Figure \ref{fig:possible_problem_with_mean_and_variance_calculation} and, therefore, is the model with the best mean-variance analysis. The Riemannian model allows us to calculate Fréchet means of the data, to calculate physically plausible geodesic deformations between the template shape and another shape, and to approximate the Riemannian distance between the template and a target shape. Hence, we have added Riemannian geometry to the latent space model.

\begin{table}[h!]
    \footnotesize
    \caption{\textbf{PGA performance.} Summary of the results of Sections \ref{sec:template_exps} and \ref{sec:template_deformation} regarding the models with Riemannian LDDMM regularization and non-Riemannian pointwise regularization. The results relate to Fig. \ref{fig:possible_problem_with_mean_and_variance_calculation}.}
    
	\centering
	\begin{tabular}{c|c||c|c}
		\toprule
		\multicolumn{1}{c@{\quad}}{\textbf{\textit{Model}}} & \multicolumn{1}{c@{\quad}}{\textbf{\textit{Dataset}}} & \multicolumn{2}{c}{\textbf{\textit{Properties}}}                 \\
		\cmidrule(r){1-1} \cmidrule{2-4}
 & & \textit{Data-consistent mean} & \textit{Proper computation variance}  \\
\cmidrule{1-4} \morecmidrules \cmidrule{1-4}
    \cmidrule{1-4}
    \textit{Non-Riemannian} & \textit{Rectangles} & \cross & \cross \\ 
    \textit{Non-Riemannian} & \textit{Liver} & \checkmark & \cross \\ 
    \cmidrule{1-4}
    \textit{Riemannian} & \textit{Rectangles} & \checkmark & \checkmark \\ 
    \textit{Riemannian} & \textit{Liver} & \checkmark & \checkmark \\ 
		\bottomrule
	\end{tabular}
\label{tab:summary_template_and_deformation_experiment}
\end{table}

\begin{table}[t!]
    \footnotesize
    \captionsetup{position=top}
    \caption{\textbf{Training and test set evaluation.} The model with the Riemannian LDDMM regularization and the model with the non-Riemannian pointwise regularization are evaluated by reconstructing the training and test sets of the rectangles (Rect.) and liver dataset. We calculate the average value and median (between brackets) of the Chamfer Distance (CD) and Earth Mover distance (EM). The CD values are of the order $10^{-4}$. The best (smallest) values are in bold.}
\begin{center}
    \subfloat[Training set evaluation]{
	\begin{tabular}{c||rr|rr|rr|rr}
		\toprule
		\multicolumn{1}{c@{\quad}}{\textbf{\textit{Model}}} & \multicolumn{8}{c}{\textbf{\textit{Dataset (metric)}}}                 \\
		\cmidrule(r){1-1} \cmidrule{2-9}
 & \multicolumn{2}{c|}{\textit{Rect. (CD)}} & \multicolumn{2}{c|}{\textit{Rect. (EM)}} & \multicolumn{2}{c|}{\textit{Liver (CD)}} & \multicolumn{2}{c}{\textit{Liver (EM)}}\\
\cmidrule{1-9} \morecmidrules \cmidrule{1-9}
  \textit{Non-Riemannian} & $\boldsymbol{0.27} $&$ (\boldsymbol{0.26})$ & $\boldsymbol{0.0187} $&$ (0.0186)$ & $\boldsymbol{1.108} $&$ (\boldsymbol{0.92})$ & $0.0274 $&$ (0.0269)$ \\
  \textit{Riemannian} &  $0.33 $&$ (0.32)$ &  $0.0188 $&$ (\boldsymbol{0.0185})$ &  $1.12 $&$ (0.95)$ &  $\boldsymbol{0.0273} $&$ ( \boldsymbol{0.0265})$\\
		\bottomrule
	\end{tabular}\label{tab:recon_metrics_train_set_rectangles}}

    \subfloat[Test set evaluation]{\label{tab:recon_metrics_test_set_rectangles}		\begin{tabular}{c||rr|rr|rr|rr}
		\toprule
		\multicolumn{1}{c@{\quad}}{\textbf{\textit{Model}}} & \multicolumn{8}{c}{\textbf{\textit{Dataset (metric)}}}                 \\
		\cmidrule(r){1-1} \cmidrule{2-9}
 & \multicolumn{2}{c|}{\textit{Rect. (CD)}} & \multicolumn{2}{c|}{\textit{Rect. (EM)}} & \multicolumn{2}{c|}{\textit{Liver (CD)}} & \multicolumn{2}{c}{\textit{Liver (EM)}}\\
\cmidrule{1-9} \morecmidrules \cmidrule{1-9}
  \textit{Non-Riemannian} & $8.20 $&$ (3.74)$ & $ 0.0326 $&$ (0.0306)$ & $\boldsymbol{5.23} $&$ (4.25)$ & $\boldsymbol{0.0369} $&$ (\boldsymbol{0.0329})$ \\
  \textit{Riemannian} &  $\boldsymbol{1.83} $&$ (\boldsymbol{1.43})$ &  $\boldsymbol{0.0245} $&$ (\boldsymbol{0.0245})$ & $5.31 $&$ (\boldsymbol{3.68})$ &  $0.0385 $&$ (0.0340)$\\
		\bottomrule
	\end{tabular}}
\end{center}
\end{table}

\subsubsection{Generalizability of shape encoding} \label{sec:generalizability_shape_encoding}
This section evaluates the reconstruction quality of the two different models by reconstructing the test sets of the rectangles and liver dataset. We use the Chamfer Distance (CD) and the Earth Mover Distance (EM) as evaluation metrics. Table \ref{tab:recon_metrics_train_set_rectangles} shows the reconstruction metrics of the training data. We see that both models perform approximately equally on the training data.

Table \ref{tab:recon_metrics_test_set_rectangles} shows the reconstruction metrics on the test sets. While the Riemannian and non-Riemannian methods on the liver dataset only penalize large deviations from the template and obtain similar reconstruction performance, the Riemannian regularization uses a rigidity prior that aligns more closely with the data on the rectangles dataset and obtains improved performance over the non-Riemannian model. This difference demonstrates the importance of a data-consistent prior on deformations to induce a meaningful prior on the latent space that enhances the model's generalization. All in all, on data with a clear Riemannian prior that not only promotes small $\norm{v}_{L^2(\Omega)}$, such as a rigidity prior, the Riemannian LDDMM regularization improves generalization. Hence, physical consistency can enhance generalization performance.

\subsubsection{Robustness to noise}
\begin{figure}[h!]
    \centering 
    \includegraphics[width=0.76\textwidth]{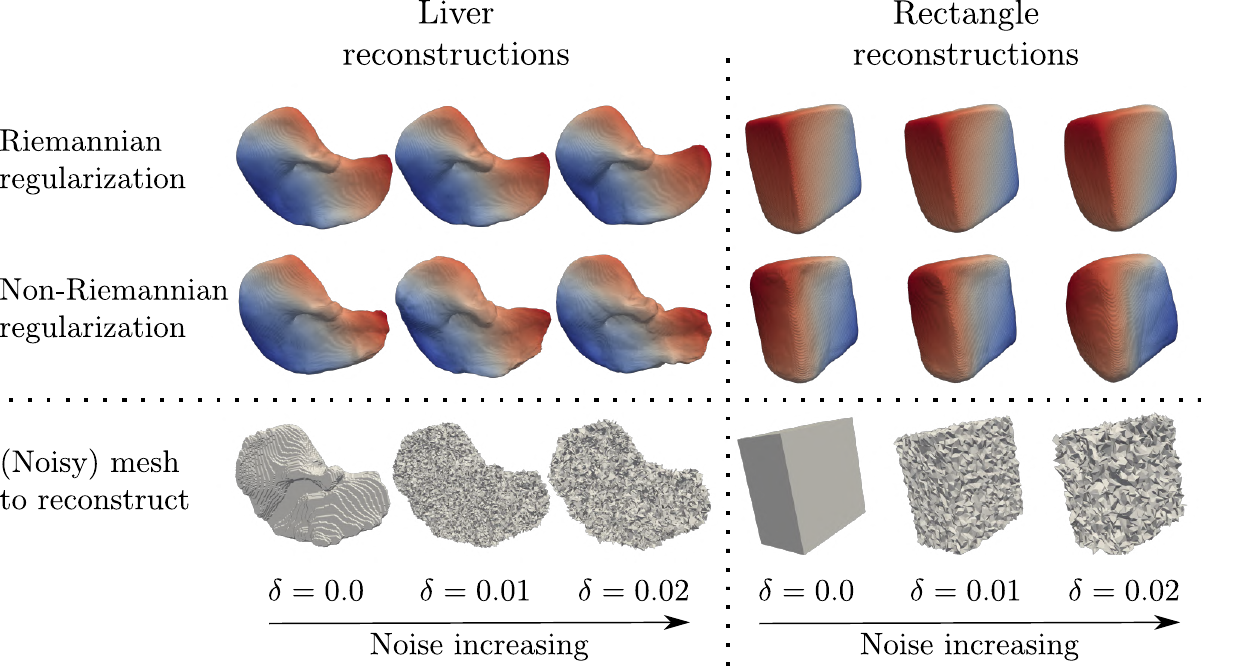}
    \caption{\textbf{Robust reconstruction.} The reconstruction of two test shapes with added zero-mean Gaussian noise for different standard deviations $\delta$. The red and blue colors represent the matching of the points to the learned template. The Riemannian regularization stabilizes the reconstruction procedure against noise.}
    \label{fig:reconstruct_noisy_shapes}
\end{figure}

In this section, we investigate how noise affects the reconstruction performance of the two models. We add random Gaussian noise with mean zero and a standard deviation $\delta$ of $0.01$ or $0.02$ to the vertices of the meshes in the test set. Subsequently, we reconstruct these noisy meshes and compare the reconstructions to the noiseless ground truth meshes. Figure \ref{fig:reconstruct_noisy_shapes} provides some of these reconstructions. In general, both models are effective in denoising the illustrated input meshes. On the rectangles, the Riemannian regularization produces good, noise-resistant reconstructions. The model with the non-Riemannian regularization generates worse reconstructions and is affected by the noise, as illustrated by the reconstructions, which gradually lose their rectangular shape when the noise increases. On the liver dataset, both models produce satisfactory denoised reconstructions. However, while the model with Riemannian regularization yields reconstructions that are approximately independent of the noise level, the model with non-Riemannian regularization results in visibly different reconstructions when the noise level varies.

To assess whether the observations concerning Figure \ref{fig:reconstruct_noisy_shapes} extend to other liver and rectangle shapes, Table \ref{tab:recon_metrics_noisy_data_rectangles} presents the average and median reconstruction errors between the reconstructed shapes and the ground truth noiseless shapes. Comparing the values to Table \ref{tab:recon_metrics_test_set_rectangles}, we notice that on the rectangles dataset, the model with Riemannian LDDMM regularization is much less sensitive to the noise than the model with the non-Riemannian regularization. On the liver dataset, the median values are relatively stable for both models. Moreover, the average reconstruction errors of both models are not influenced much by a noise level of $0.01$. However, when increasing the standard deviation to $0.02$, the non-Riemannian model attains a much bigger average reconstruction error. In contrast, the Riemannian model attains a similar average error to the scenario with a $0.01$ noise level.

\begin{table}[t!]
    \footnotesize
	\caption{\textbf{Evaluating robustness.} The models with the non-Riemannian pointwise and the Riemannian LDDMM regularization are compared by reconstructing noisy test sets ($\delta=0.01$ and $\delta=0.02$) of the rectangles (Rect.) and liver dataset. We calculate the average value and median (between brackets) of the Chamfer Distance (CD) and Earth Mover distance (EM). The values are obtained by comparing the noiseless shapes to the reconstructions of their noisy versions. The CD values are of order $10^{-4}$. The two smallest values are in bold, showing the increased robustness of the Riemannian approach. 
    }
    
	\centering
	\begin{tabular}{c|c||rr|rr|rr|rr}
		\toprule
		\multicolumn{1}{c@{\quad}}{\textbf{\textit{Model}}} & \multicolumn{1}{c@{\quad}}{$\boldsymbol{\delta}$} & \multicolumn{8}{c}{\textbf{\textit{Dataset (metric)}}}                 \\
		\cmidrule(r){1-2} \cmidrule(r){3-10} 
 & & \multicolumn{2}{c|}{\textit{Rect. (CD)}} & \multicolumn{2}{c|}{\textit{Rect. (EM)}} & \multicolumn{2}{c|}{\textit{Liver (CD)}} & \multicolumn{2}{c}{\textit{Liver (EM)}}\\
\cmidrule{1-10} \morecmidrules \cmidrule{1-10}
  \textit{Non-Riemannian} & $0.01$ & $39.37 $&$(4.48)$ & $0.0409  $&$(0.0310)$ & $6.57 $&$ (\boldsymbol{3.97})$ & $\boldsymbol{0.0405} $&$ (0.0369)$ \\ 
  \textit{Non-Riemannian} & $0.02$ & $54.82 $&$(8.17)$ & $0.0502  $&$(0.0367)$ & $10.49 $&$ (5.48)$ & $0.0451 $&$ (0.0387)$ \\ 
  \textit{Riemannian} & $0.01$  & $\boldsymbol{2.01} $&$ (\boldsymbol{1.72})$ & $\boldsymbol{0.0242}  $&$(\boldsymbol{0.0252})$ & $\boldsymbol{5.95}  $&$(\boldsymbol{3.82})$ & $\boldsymbol{0.0390}  $&$(\boldsymbol{0.0322})$ \\ 
  \textit{Riemannian} & $0.02$  & $\boldsymbol{4.06} $&$ (\boldsymbol{3.43})$ & $\boldsymbol{0.0275}  $&$(\boldsymbol{0.0276})$ & $\boldsymbol{6.53}  $&$(4.89)$ & $0.0405  $&$(\boldsymbol{0.0365})$ \\ 
		\bottomrule
	\end{tabular}
\label{tab:recon_metrics_noisy_data_rectangles}
\end{table}

The latter observation is related to Figure \ref{fig:reconstruct_noisy_shapes} where the liver reconstructions obtained via the non-Riemannian model depend on the noise level. As discussed in Section \ref{sec:template_deformation}, the non-Riemannian regularization encourages a rapid transition to the target shape followed by smaller deformations for fine-tuning. Such large displacements can cause sensitivity to small changes and might yield worse or different reconstructions. Disallowing such large displacements via the Riemannian velocity field regularization stabilizes the problem. 

In summary, the physical consistency obtained by the Riemannian regularization improves robustness. The improved robustness is largest on data with a clear Riemannian prior that not only promotes small $\norm{v}_{L^2(\Omega)}$, such as the rigidity prior of the rectangles dataset.

\subsection{INRs for LDDMM statistical latent modeling}
\label{sec:INR_for_stat_model}
To illustrate the benefit of INRs for topology-preserving latent modeling, we use the liver dataset and compare our method to the diffeomorphic autoencoder (DAE) \cite{bone2019learning}, which works directly on meshes. This method parameterizes a latent space of discrete velocity vector fields on a rectangular grid and obtains a time-dependent discrete velocity vector field via the linear interpolation from the origin to some other latent vector. This time-dependent velocity vector field deforms a learnable template mesh to obtain shape reconstructions. Image representations like signed distance functions for contours or shapes are not always available and non-trivial to obtain from meshes or point clouds. Hence, we can not easily compare our more general framework to LDDMM-based statistical latent modeling methods, like Hinkle et al.\ \cite{hinkle2018diffeomorphic}, that have this requirement. 

\begin{figure}[h!]
     \centering
    \subfloat[Original data]{\includegraphics[width=0.3\textwidth]{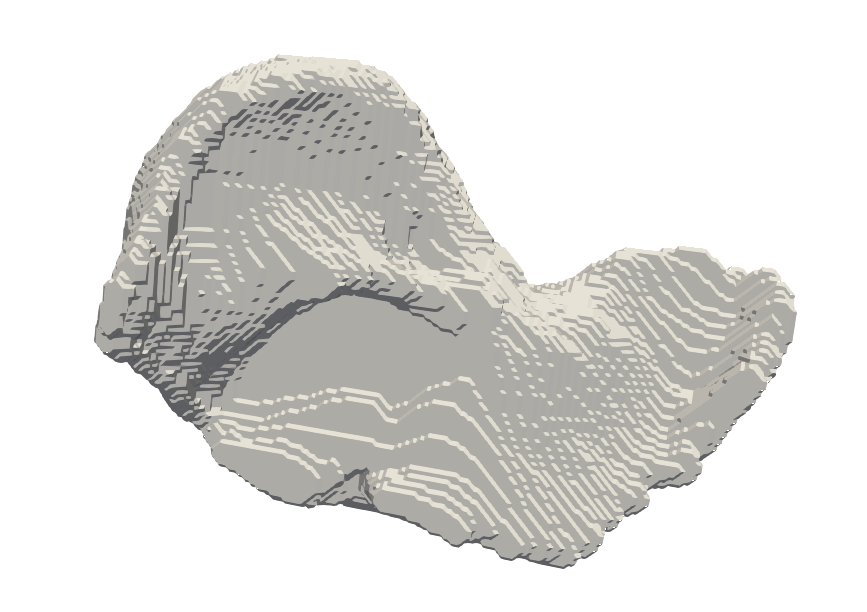}}
    \hfill
    \subfloat[Downsampled non-smooth data]{\includegraphics[width=0.3\textwidth]{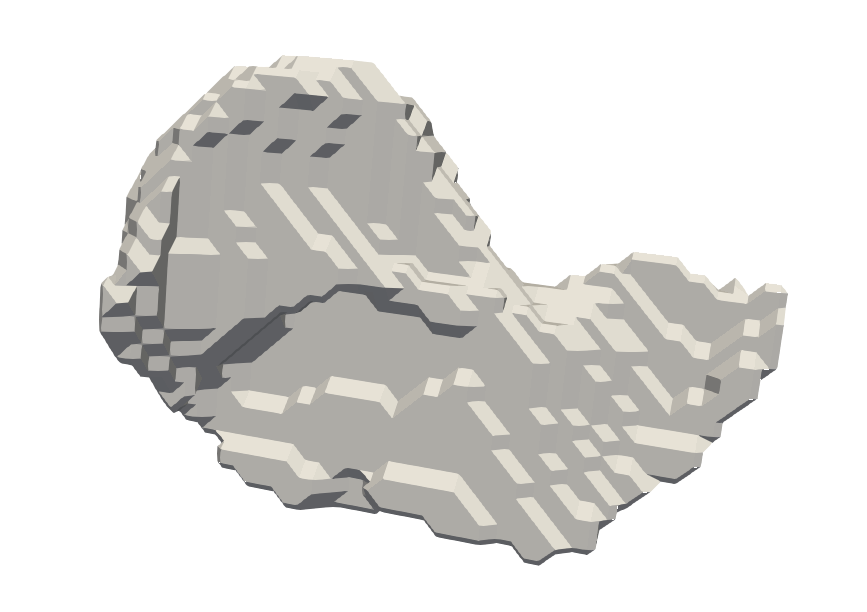}}
    \hfill
    \subfloat[Downsampled smooth data]{\includegraphics[width=0.3\textwidth]{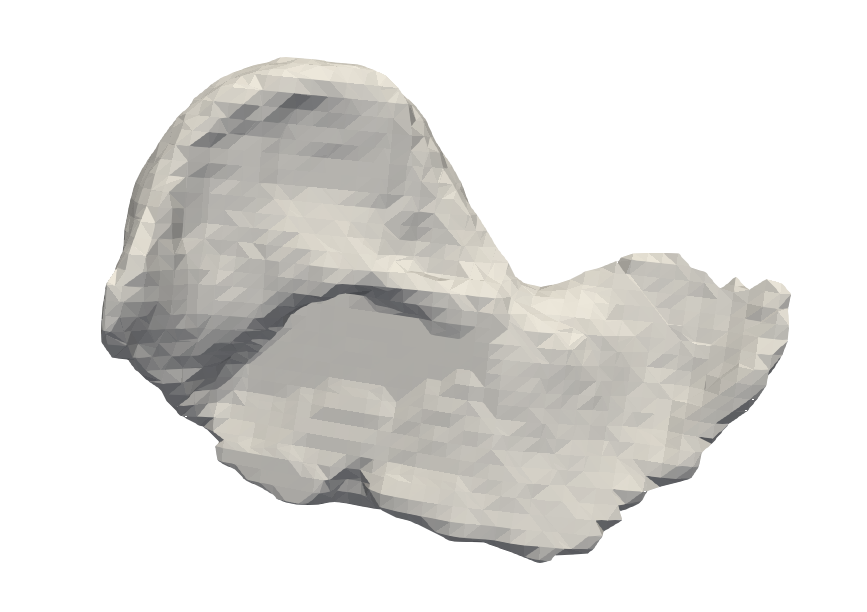}}
        \caption{\textbf{Liver training data.} From left to right: the original mesh data, the downsampled version, and the downsampled smooth version.}
        \label{fig:non_smooth_vs_smooth_and_low_res_vs_high_res}
\end{figure}

First, we train on our original mesh data containing many vertices but without a smooth surface. Subsequently, we do the same experiment for downsampled meshes with a much smaller amount of vertices: one dataset with non-smooth meshes and one dataset with smoother meshes. The three mentioned datasets are depicted in Figure \ref{fig:non_smooth_vs_smooth_and_low_res_vs_high_res}. To obtain the downsampled non-smooth meshes, we take the original mesh data, calculate a 64 x 64 x 64 occupancy grid for the meshes, and subsequently perform marching cubes \cite{lorensen1987marching} on the occupancy grid. To obtain a smooth version of the meshes, we also calculate a signed distance grid and apply marching cubes to it. 

We train the DAE model only on the downsampled data as the model failed to converge to a proper solution on the original data. On the other hand, we only train our model on the original data and the downsampled non-smooth data. The reason is that this is noisier data and we shall see that our algorithm is agnostic to such noise.

\begin{figure}[t!]
     \centering
    \subfloat[RDA-INR non-smooth data]{\includegraphics[width=0.3\textwidth]{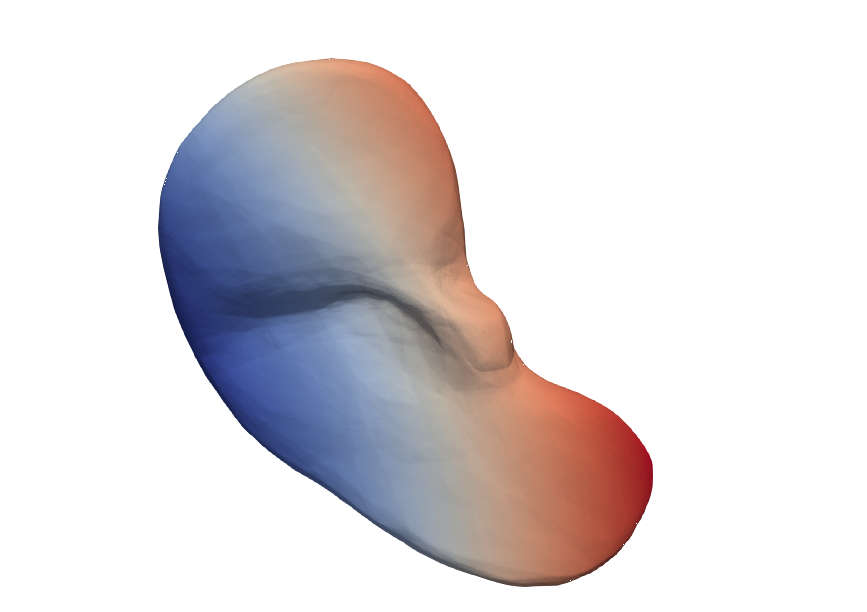}}
    \hfill
        \subfloat[DAE non-smooth data]{\includegraphics[width=0.3\textwidth]{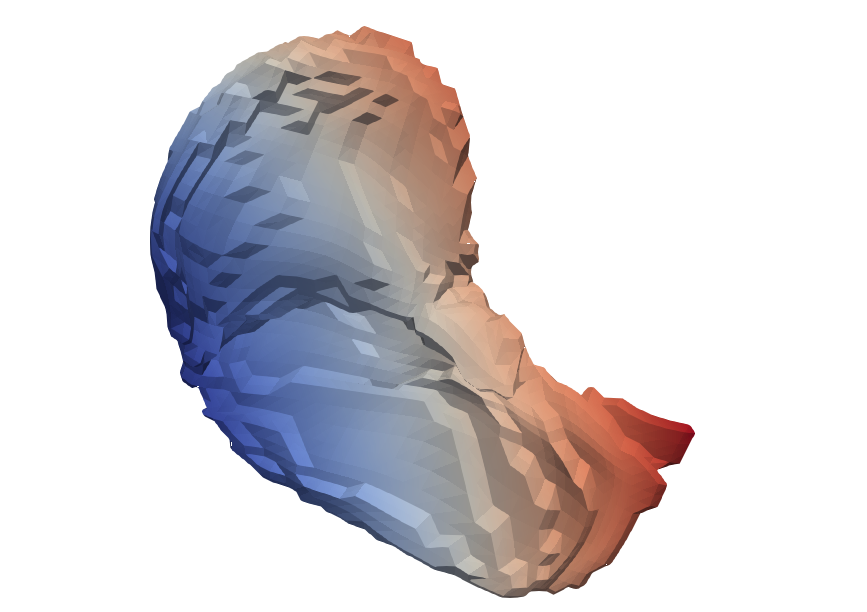}}
    \hfill
        \subfloat[DAE smooth data]{\includegraphics[width=0.3\textwidth]{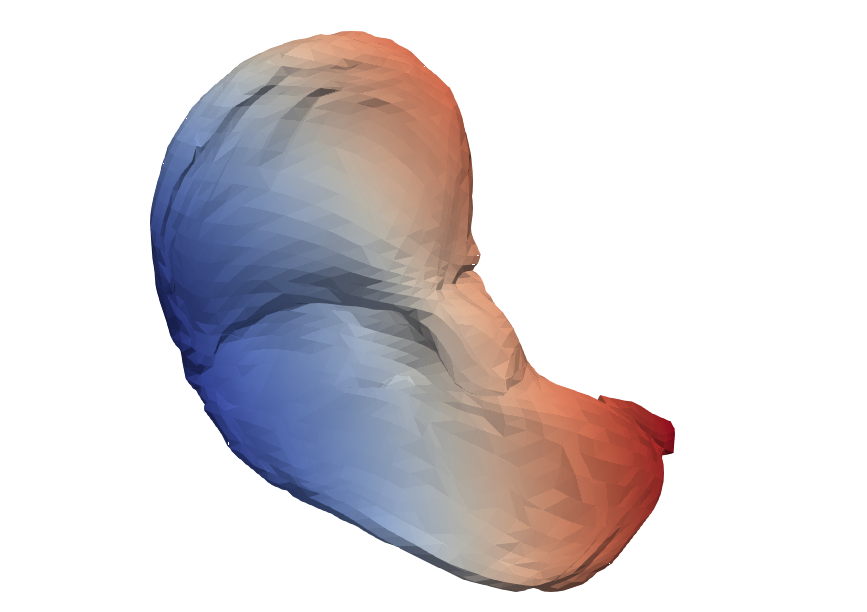}}
        \caption{\textbf{Learned templates.} The learned templates of RDA-INR and DAE on the downsampled datasets in Figure \ref{fig:non_smooth_vs_smooth_and_low_res_vs_high_res}. RDA-INR is robust against limited data quality (subsampling artifacts, discretization). }
        \label{fig:templates_DAE}
\end{figure}

First, we evaluate the learned templates trained on the downsampled data. These learned templates can be found in Figure \ref{fig:templates_DAE}. The template learned by our model resembles the template that is learned using the original data, which is depicted in Figure \ref{fig:templates_subfigure_a}. Hence, even with less detailed meshes we can learn a high-resolution template. Furthermore, while the DAE template found using the non-smooth data looks non-smooth, the template found with the smooth data looks quite smooth. The latter looks similar to the liver template found with the RDA-INR model. However, the template found by DAE is less smooth than the template found by our model. The main reason for this is the smooth structure of the INR parameterizing the Fréchet mean. In DAE the Fréchet mean is calculated by Deformetrica \cite{bone2018deformetrica}. For Deformetrica we take a single sample from the dataset as initialization of the Fréchet mean. Hence, we inherit the structure of this initialization, possibly causing a problem with the template.

\begin{figure}[h!]
     \centering

        \subfloat[RDA-INR fully detailed original data]{\includegraphics[width=0.2\textwidth]{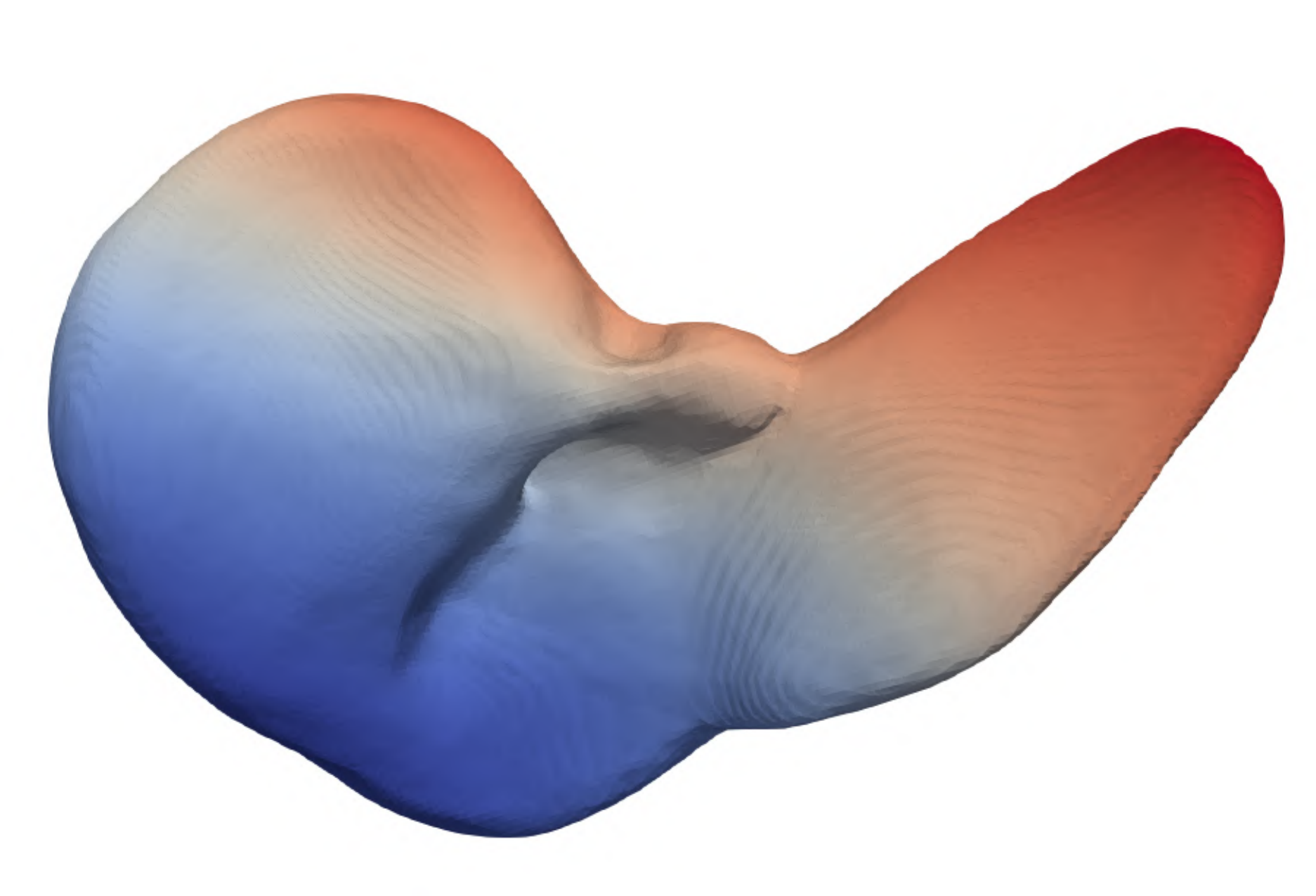}}
        \hfill
        \subfloat[RDA-INR downsampled non-smooth data]{\includegraphics[width=0.2\textwidth]{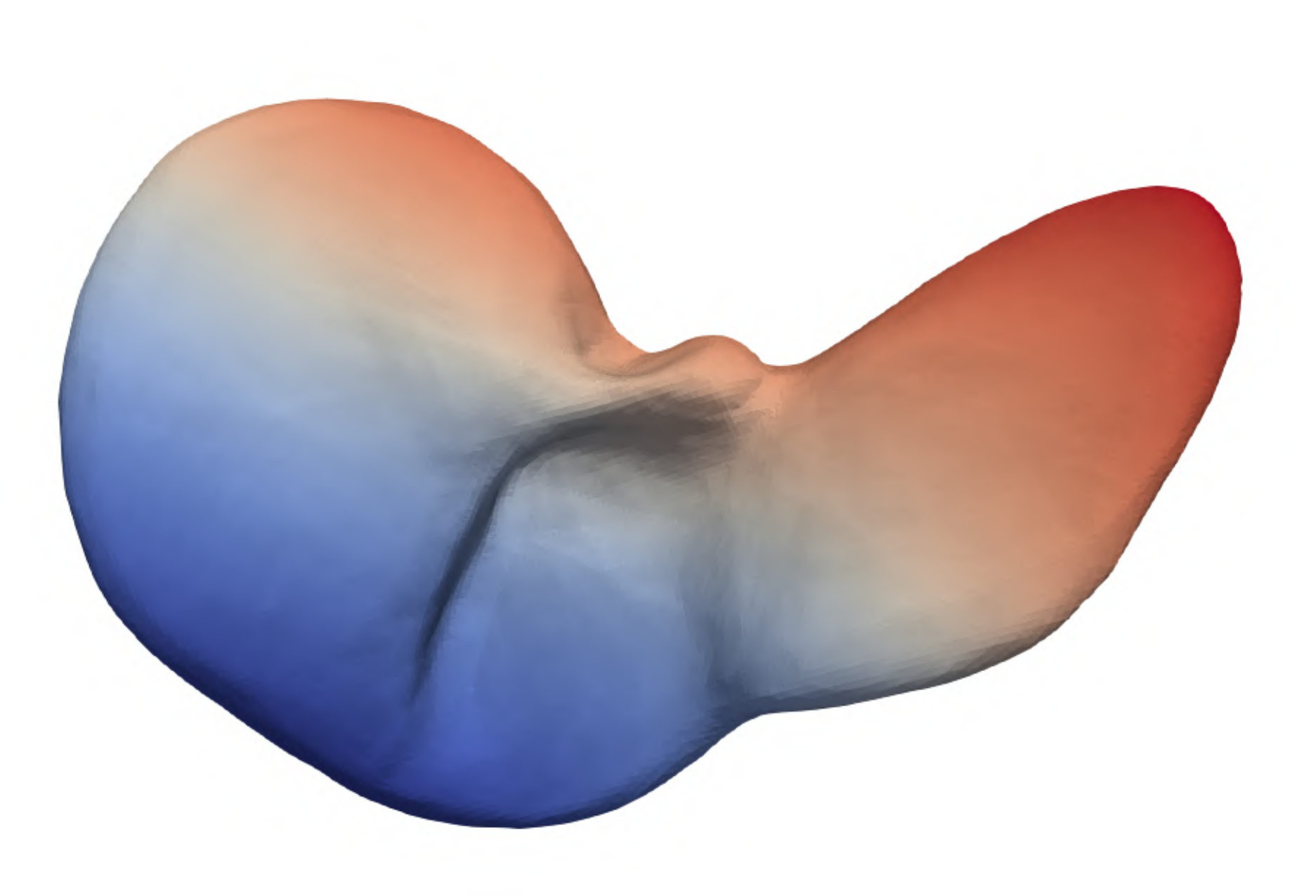}}
        \hfill
        \subfloat[DAE downsampled non-smooth data]{\includegraphics[width=0.2\textwidth]{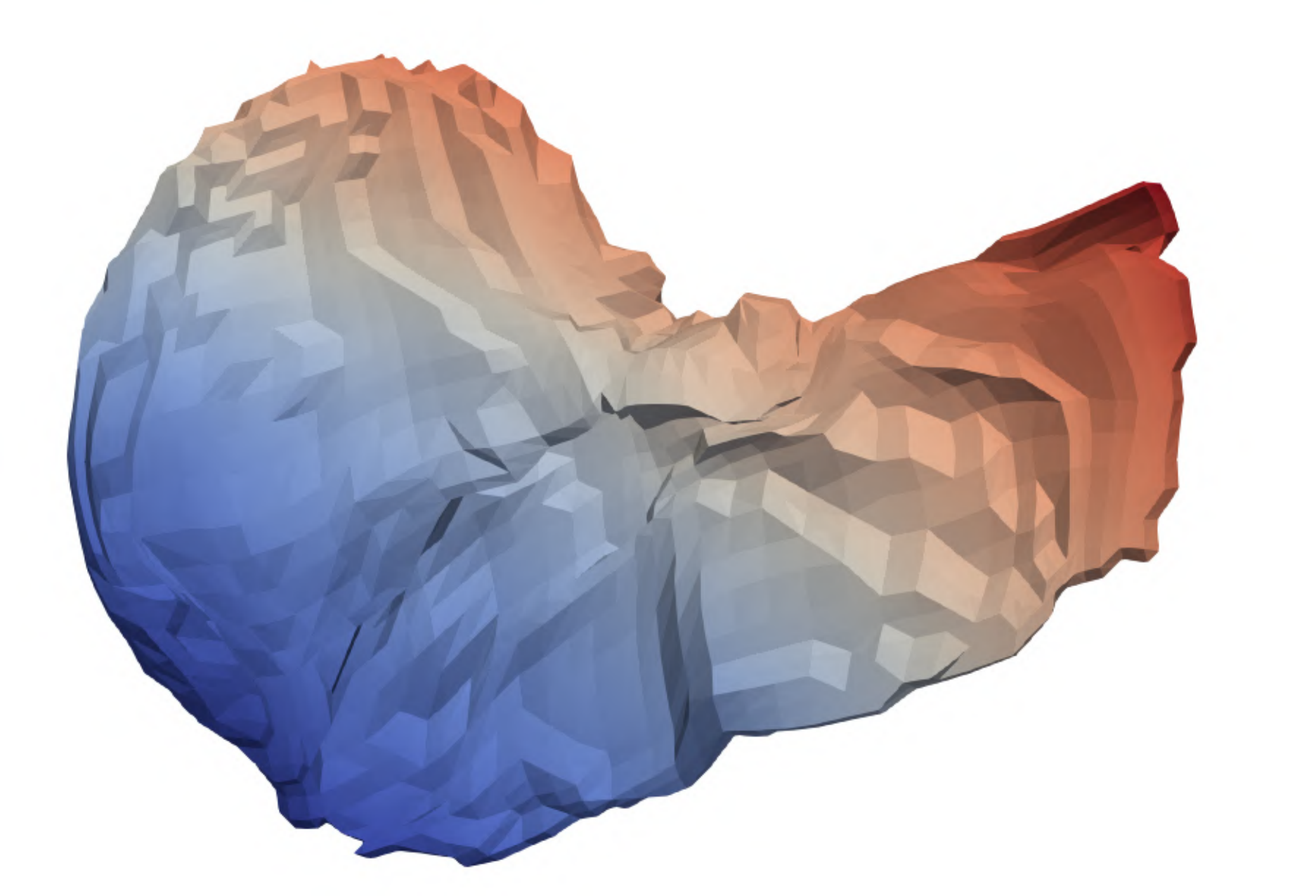}}
        \hfill
        \subfloat[DAE downsampled smooth data]{\includegraphics[width=0.2\textwidth]{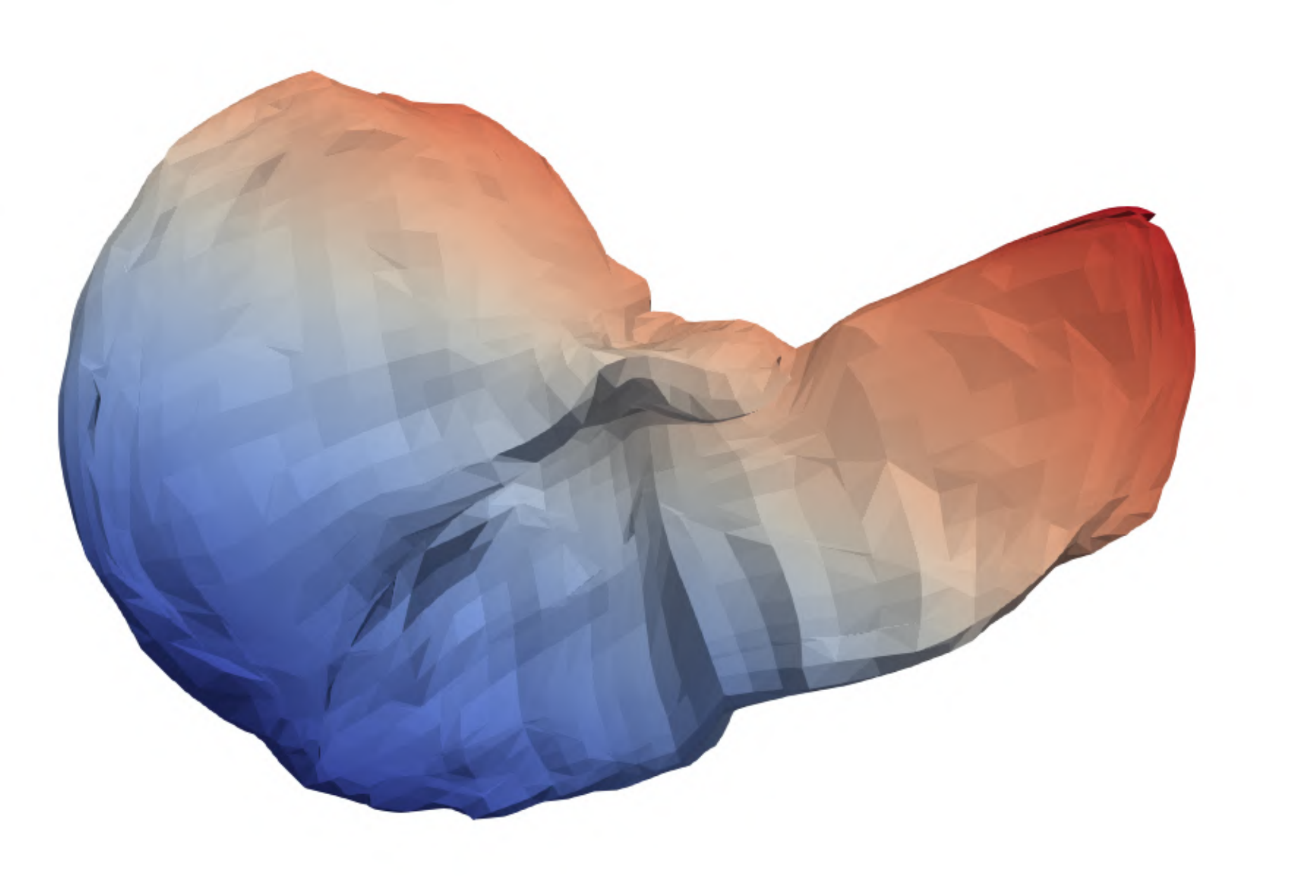}}
        \caption{\textbf{Super resolution reconstruction.} Test sample reconstruction on different variants of the original liver data. RDA-INR achieves similar reconstructions independent of resolution.}
        \label{fig:reconstructions_benefit_INR_experiment}
\end{figure}

The effect of the templates on the reconstructions of test samples is shown in Figure \ref{fig:reconstructions_benefit_INR_experiment}. We see that the reconstructions are heavily affected by the quality of the templates. On the other hand, our model does not seem affected by the resolution of the dataset as the reconstructions of the model trained on the original data and the downsampled non-smooth data are very similar. To strengthen these observations, we quantitatively assess reconstruction performance. We compare the reconstructed meshes to the ground-truth mesh with many vertices via the Chamfer Distance and the Earth Mover Distance. These metrics are calculated between the reconstructed mesh and many points sampled from the original mesh. The results are presented in Table \ref{tab:recon_metrics_super_resolution_task}. We can see that our model outperforms the DAE method and that our model does not deteriorate much when using the downsampled data. 

\begin{table}[h!]
    \footnotesize
	\caption{\textbf{Quantitative reconstruction performance.} The table presents the average values and medians (between brackets) of the Chamfer Distance (CD) and Earth Mover distance (EM). The values are obtained by comparing the original data to the reconstructions obtained using the dataset the model is trained on. These values are calculated for the shapes in the train and test splits of the liver dataset. The CD values are of the order $10^{-4}$. The best (smallest) values for the downsampled data are in bold.}
    
	\centering
	\begin{tabular}{l @{\hspace{0.7\tabcolsep}} l||rr|rr|rr|rr}
		\toprule
		\multicolumn{2}{c@{\quad}}{\textbf{\textit{Model (Dataset)}}} & \multicolumn{8}{c}{\textbf{\textit{Data split (metric)}}}                 \\
		\cmidrule(r){1-2} \cmidrule{3-10} 
 & & \multicolumn{2}{c|}{\textit{Train (CD)}} & \multicolumn{2}{c|}{\textit{Train (EM)}} & \multicolumn{2}{c|}{\textit{Test (CD)}} & \multicolumn{2}{c}{\textit{Test (EM)}}\\
\cmidrule{1-10} \morecmidrules \cmidrule{1-10}
  \textit{RDA-INR} & \textit{(Original data)} & $1.12 $&$ (0.95)$ &  $0.0273 $&$ ( 0.0265)$ & $5.31 $&$ (3.68)$ &  $0.0385 $&$ (0.0340)$ \\ 
  \cmidrule{1-10}
  \textit{DAE} & \textit{(Non-smooth)} & $4.47 $&$(4.01)$ & $0.0350 $&$(0.0339)$ & $13.52 $&$ (10.22)$ & $ 0.0479 $&$ (0.0443)$ \\ 
  \textit{DAE} & \textit{(Smooth)} & $4.12 $&$(3.54)$ & $0.0344 $&$(0.0330)$ & $14.46 $&$ (11.10)$ & $0.0492 $&$ (0.0464)$ \\ 
  \textit{RDA-INR} & \textit{(Non-smooth)} & $\boldsymbol{1.61} $&$ (\boldsymbol{1.45})$ & $\boldsymbol{0.0288}  $&$(\boldsymbol{0.0282})$ & $\boldsymbol{6.34}  $&$(\boldsymbol{4.17})$ & $\boldsymbol{0.0402}  $&$(\boldsymbol{0.0351})$ \\ 
		\bottomrule
	\end{tabular}
\label{tab:recon_metrics_super_resolution_task}
\end{table}

\section{Conclusion and future work}\label{sec:conclusion}
Several works recently developed neural network models regarding diffeomorphic registration, such as models for pairwise and groupwise registration, atlas building, and data variability modeling. For instance, there exist LDDMM-based statistical latent modeling approaches, which have the limitation that they are not resolution independent. Moreover, resolution-independent neural network algorithms exist for joint shape encoding and groupwise registration. However, they do not use the Riemannian geometry of shape space, which is a crucial component of LDDMM PGA. Consequently, these latent space models do not provide insights about shape Fréchet means, geodesics, Riemannian distances between shapes, and data variance. 

We have addressed the aforementioned two limitations to highlight the importance of resolution independence and physical consistency. Specifically, we addressed these issues in the setting of shapes represented by point clouds or meshes. In this setting, we presented a resolution-independent latent model based on INRs and inspired by LDDMM-based PGA. The main ingredient in this deformable template model is the Riemannian regularization on the neural network that deforms the template. Furthermore, to the best of our knowledge, our latent model is the first in the registration literature using image-based (implicit neural) data representations for dealing directly with point cloud or mesh data.

We use the model to provide a study on the benefits of resolution independence and LDDMM physical consistency. First, we discussed the importance of LDDMM physical consistency by comparing our model to the model with a non-Riemannian pointwise regularization on the deformation. We show that the Riemannian regularization is necessary for the model to perform a proper mean-variance analysis. In other words, our model allows calculating a Fréchet mean of the data, obtaining geodesics and approximating the distance between the template and another object, and estimating the data variance. Furthermore, we demonstrate that the Riemannian regularization can improve the reconstruction of an object and the stability of the reconstruction procedure. In other words, the Riemannian regularization induces a prior that enables us to find more stable factors of variation. Finally, we assess the importance of resolution-independence by comparing our model to another neural network model inspired by LDDMM PGA. We demonstrated improved template learning and improved reconstructions. 

In summary, we show how shape and image analysis, Riemannian geometry, and deep learning can be connected. This connection paves the way for more research into how these different disciplines can reinforce each other. 

\subsection{Future work}
Our learned template deformations constitute geodesics between the template and the reconstructed objects. This makes it possible to obtain template deformations that fit a prior and to reconstruct objects using the end point of such a deformation. However, for some objects in the geodesic, there might not exist a latent code and corresponding reconstruction. A future extension of our framework could be to add this missing geodesic consistency to enforce structure on the latent space. This structure increases interpretability as generally changes in latent space have an a-priori unclear effect on the reconstruction. To add the geodesic consistency, we might employ the velocity vector field parameterization from Lüdke et al.\ \cite{ludke2022landmark}.

Moreover, a related future impact of the model could be its use for generative modeling. Based on the Riemannian distance, Fréchet means, and geodesics, we can design a Riemannian probability distribution on the latent space. This reveals a part of the latent space's structure, making it more interpretable and easier to navigate.

Our model is not only applicable to point cloud and mesh data but also to images. In future work, we would like to apply our model to image data towards understanding the importance of data representation in this framework.

\section*{Acknowledgments}
The authors would like to thank the associate editor and the two anonymous referees for their time and effort to
review our manuscript thoroughly and for their suggestions on strengthening the reported work.

\bibliographystyle{siamplain}
\bibliography{references}

\newpage
\appendix

\section{Riemannian geometry}\label{app:riem_geom_basics}
In this section, we briefly present a high-level overview of some key concepts of differential geometry and Riemannian geometry. For more in-depth information, we refer to \cite{jost2008riemannian}. 

Manifolds are the main object in Riemannian geometry and differential geometry. Intuitively, a $d$-dimensional manifold $M$ is a set that locally looks like $\mathbb{R}^d$. An example of a manifold is the sphere. Moreover, a submanifold $N$ of $M$ is a subset of $M$ that is also a manifold.

To define differentiability on manifolds and submanifolds, we need a differentiable manifold. An important notion for differentiable manifolds is the tangent space $T_p M$ at a point $p \in M$, which can be thought of as the tangent plane to the manifold at $p$. More formally, the tangent space can be defined as:
\begin{definition}[Tangent space]
Assume we have a smooth curve $\gamma: \mathbb{R} \rightarrow M$ on a manifold $M$ with $\gamma(0)=p$. Define the directional derivative operator at $p$ along $\gamma$ as
\begin{align*}
    X_{\gamma, p}:C^\infty(M) & \rightarrow \mathbb{R} \\
    f & \rightarrow (f \circ \gamma)'(0),
\end{align*}
where $C^\infty(M)$ denotes the set of smooth scalar functions on $M$. Then the tangent space $T_p M$ is defined as $T_p M := \{X_{\gamma,p} \mid \gamma(0)=p, \gamma \text{ smooth} \}$.
\end{definition}

These tangent spaces can be used to define shortest paths on manifolds. For defining shortest paths, we need a Riemannian manifold:
\begin{definition}[Riemannian manifold]
    Let $M$ be a differentiable manifold. Define a Riemannian metric $g$ as a smoothly varying metric tensor field. In other words, for each $p\in M$, we have an inner product $g_p:T_pM \times T_pM \rightarrow \mathbb{R}$ on the tangent space $T_p M$. The pair $(M, g)$ is called a Riemannian manifold. 
\end{definition}

We note that submanifolds $N$ inherit the differential structure and the Riemannian metric structure of $M$. Using the metric structure, we define shortest paths between $p$ and $q$ as minimizers of:
\begin{equation}
\begin{aligned}
        d_M(p,q) = \min_{\gamma} & \int_0^1 \sqrt{g_{\gamma(t)}(\dot{\gamma}(t), \dot{\gamma}(t))} \mathrm{d}t \\
        \textrm{s.t.} & \quad \gamma(0)=p, \gamma(1)=q,
\end{aligned}
\label{eq:riem_dist}
\end{equation}
where $\dot{\gamma}(t)$ is the tangent vector at $\gamma(t)$ generated by $\gamma$. In this case, Equation \eqref{eq:riem_dist} defines a Riemannian distance on $M$ and the minimizer $\gamma^*$ is called a (Riemannian) geodesic. 
\begin{remark}
    In case $M = \mathbb{R}^d$ and $g_p(a,b) = \langle a, b \rangle_2$, we have $d_M(p,q) = \norm{p - q}_2$ and $\gamma^*(t) = p + t (q - p)$. 
\end{remark}

Given a geodesically complete manifold, for any point $p \in M$ and tangent vector $\dot{p} \in T_p M$, there exists a unique geodesic $\gamma$ with $\gamma(0) = p$ and $\dot{\gamma}(0) = \dot{p}$. The unique solution at $t=1$ is given by the exponential map $\text{exp}_p(\dot{p}):=\gamma(1)$. The exponential map allows mapping tangent vectors in $T_p M$ to points on the manifold. Consequently, we can execute several manifold operations on a tangent space instead of on the manifold. For instance, the exponential map is used in PGA to obtain a manifold version of PCA.

Finally, the Riemannian distance allows us to define manifold extensions of means in vector spaces and allows us to define a specific type of submanifold:
\begin{definition}[Fréchet mean]
    Let $\rho$ be a probability distribution on $M$. The Fréchet mean $\mu$ is defined as
    \begin{equation*}
        \mu = \argmin_{q \in M} \int_M d_M^2(p,q) \mathrm{d}\rho(p).
    \end{equation*}
    If we only have a finite sample $\{p_i\}_{i=1}^N$ with $p_i \in M$, the Fréchet mean $\mu$ is defined as
    \begin{equation*}
        \mu = \argmin_{q \in M} \frac{1}{N}\sum_{i=1}^N d_M^2(p_i, q).
    \end{equation*}
\end{definition}
\begin{definition}[Geodesic submanifold]
   A geodesic submanifold of a Riemannian manifold $M$ is a submanifold $N$ such that $\forall p \in N$, all geodesics of $N$ passing through $p$ are also geodesics of $M$. 
\end{definition}

\newpage

\section{Additional experiments}
\subsection{Occupancy data versus point cloud data} \label{app:occnet_vs_IGR}
In the main text, we discuss two data fidelity terms that can be used in our implicit encoding model. One approach uses occupancy functions with a binary cross entropy data fidelity term, while the other approach uses point cloud data. To showcase the difference between the two approaches, we perform an experiment on the rectangle data as used in the numerical results section. The exact same training parameters are used.

\begin{figure}[h!]
     \centering
     \subfloat[Occupancy data.]{\includegraphics[width=0.34\textwidth]{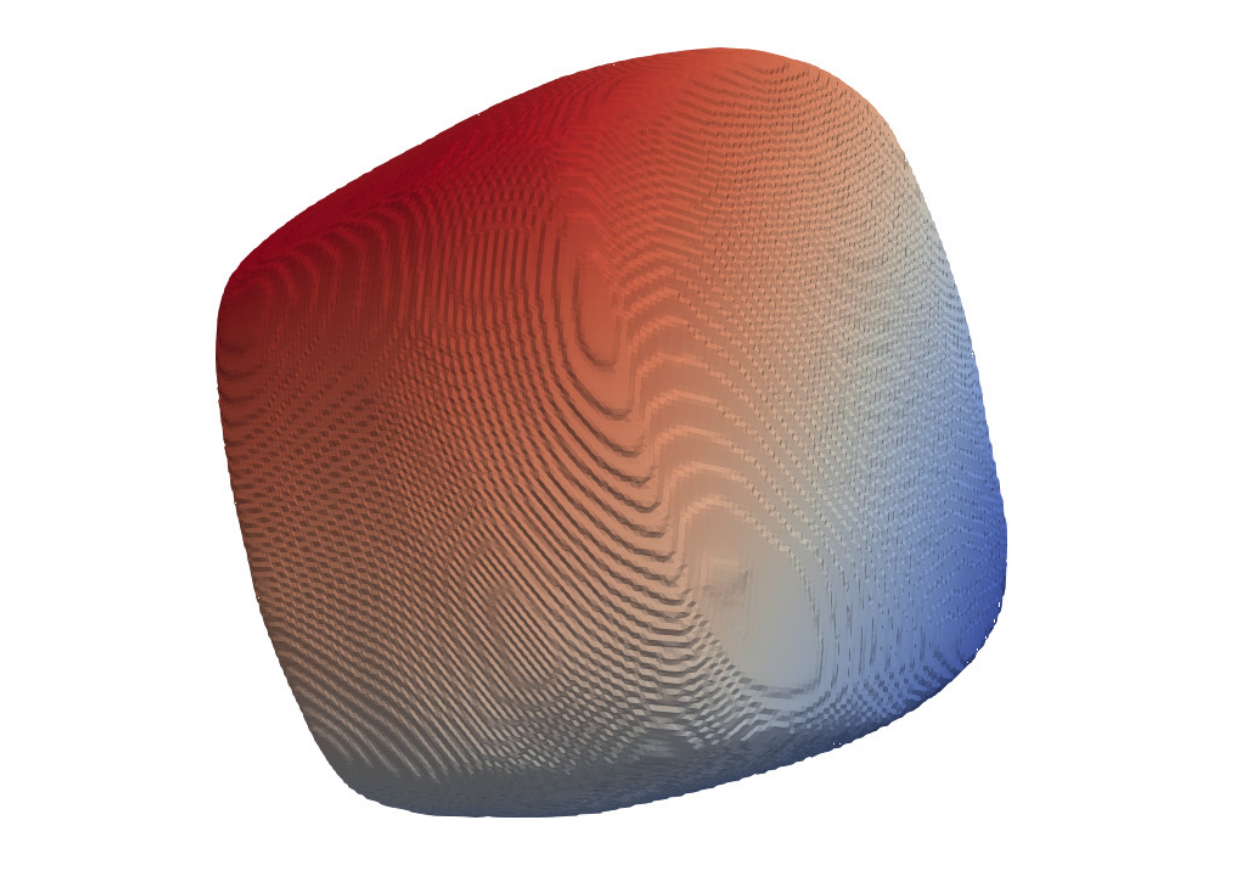}}
     \hfill
      \subfloat[Point cloud data.]{\includegraphics[width=0.34\textwidth]{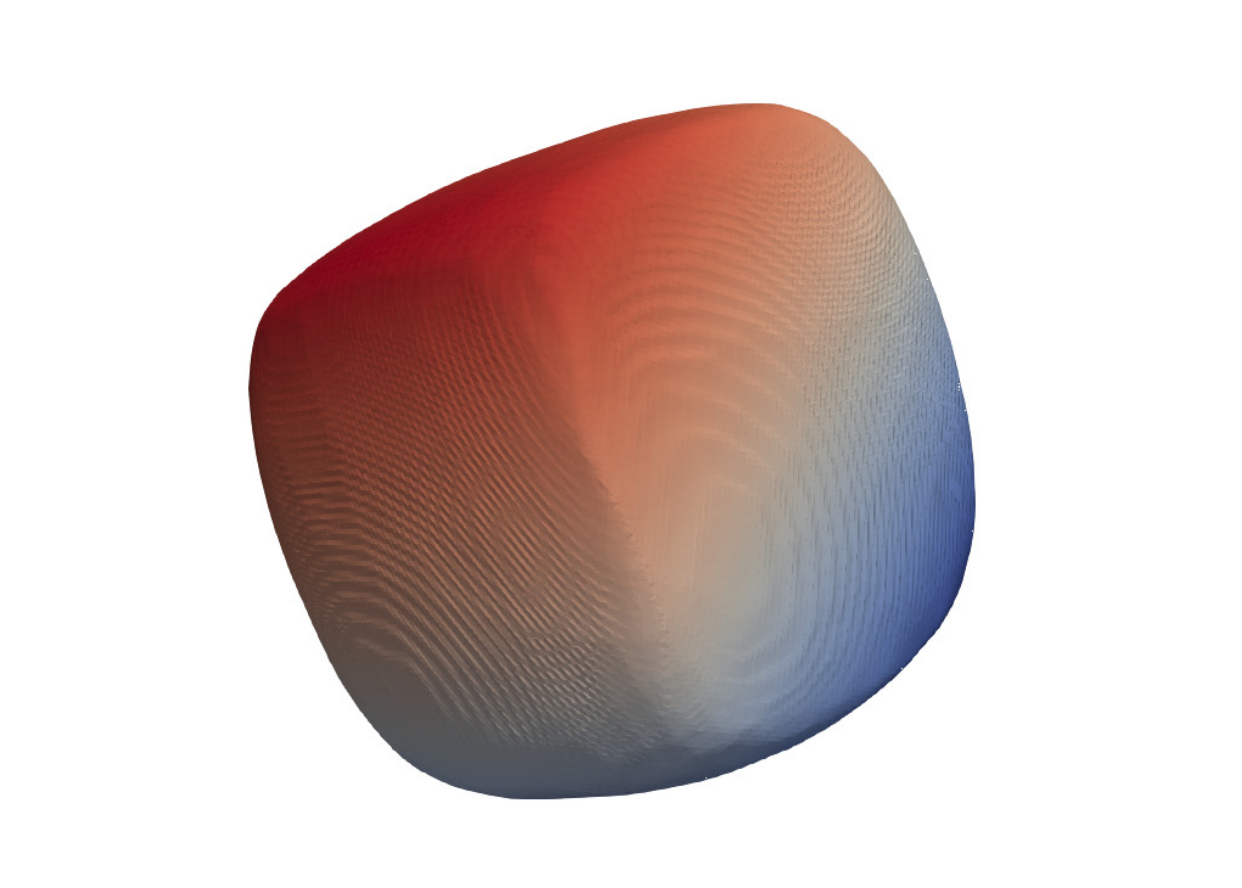}}
        \caption{\textbf{Learned templates.} The template of the model learned with occupancy functions (left) and the model learned with the point cloud data (right). Both templates resemble the rectangles data.}
        \label{fig:templates_occ_vs_point cloud}
\end{figure}

\begin{figure}[h!]
     \centering
     \subfloat[Occupancy data.]{\includegraphics[width=0.34\textwidth]{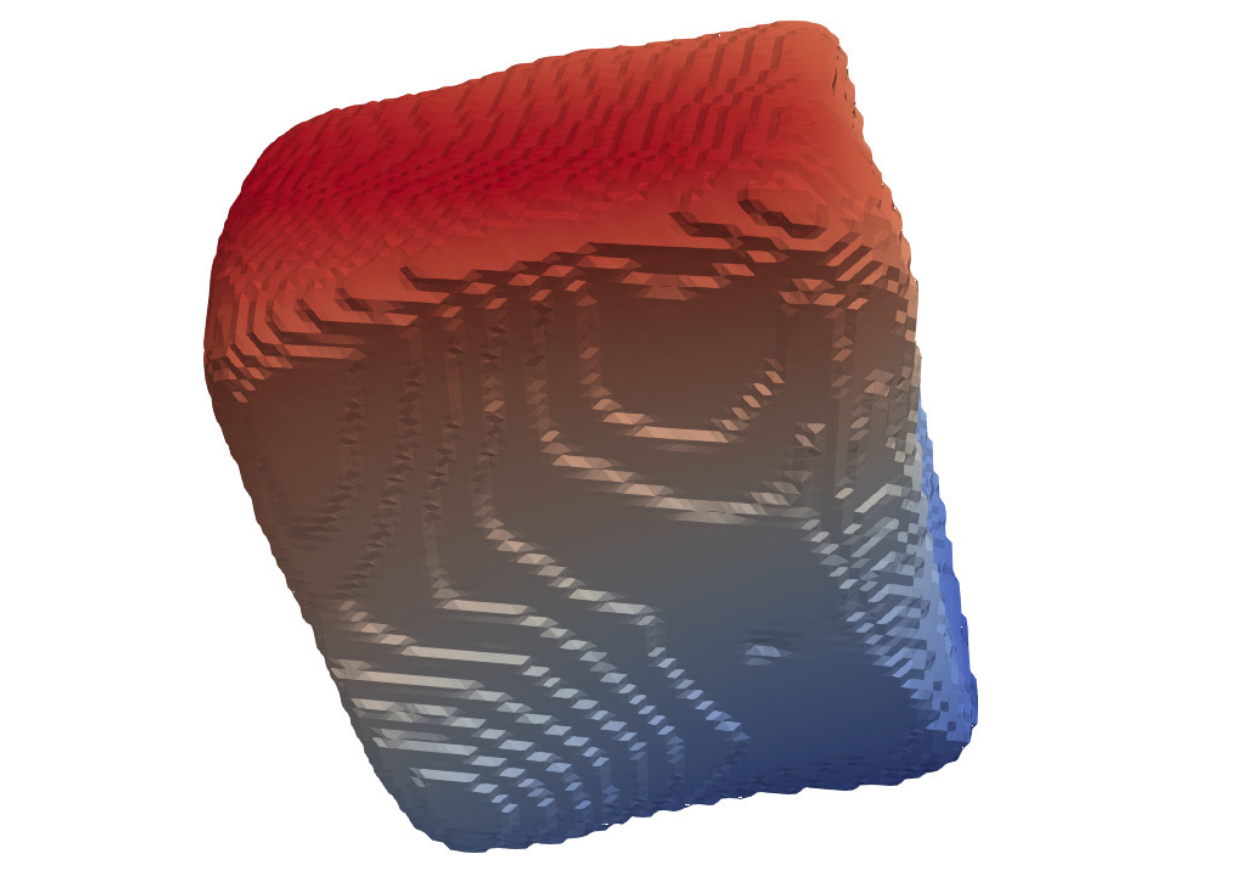}}
     \hfill
      \subfloat[Point cloud data.]{\includegraphics[width=0.34\textwidth]{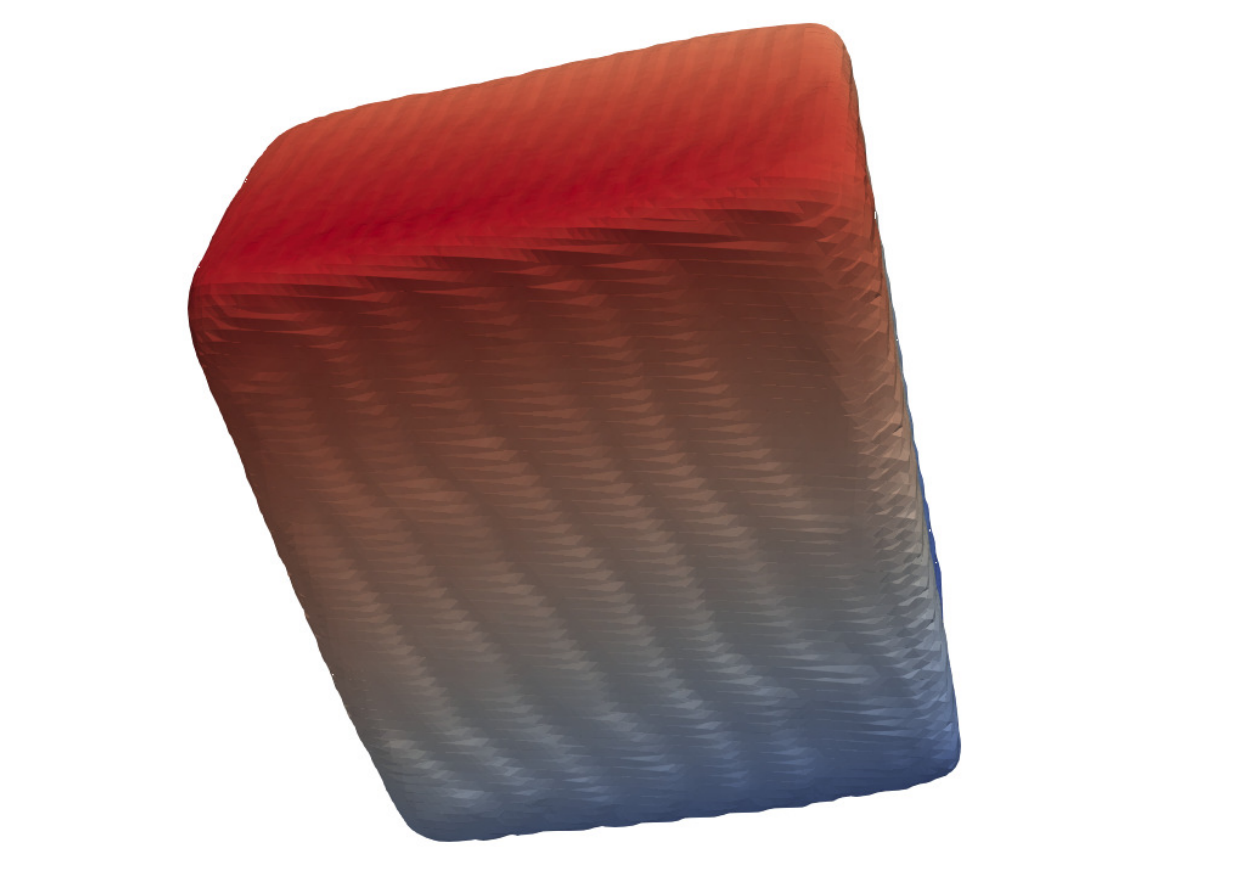}}  
        \caption{\textbf{Training shape reconstruction.} The reconstructions of a particular training shape when using the model learned with occupancy functions (left) and the model learned with the point cloud data (right). The model with the point cloud data obtains a higher quality reconstruction.}
        \label{fig:reconstructions_occ_vs_point cloud}
\end{figure}

In Figures \ref{fig:templates_occ_vs_point cloud} and \ref{fig:reconstructions_occ_vs_point cloud}, we see the learned templates and an example of a reconstructed training shape, respectively. 

Figure \ref{fig:templates_occ_vs_point cloud} shows that both templates resemble a square. Figure \ref{fig:reconstructions_occ_vs_point cloud} demonstrates that the reconstructions with the occupancy data are worse than the reconstructions with the point cloud data. One possible explanation is that the point cloud loss function encourages the shape's points to lie on the zero level set of the implicit representation. When using occupancy values, we focus more on the domain around the shape and train on uniformly sampled occupancy values to regress the occupancy function. The focus on the shape itself makes it possible to better reconstruct its details. As the point cloud method yields better reconstructions, we use this method for the numerical experiments. 

\subsection{Comparison RDA-INR and NDF}\label{app:comparison_RDA-INR_NDF}
In the main manuscript, we note that our model can be interpreted as a modification of Neural Diffeomorphic Flow (NDF) \cite{sun2022topology}. For instance, we modify the data attachment such that we can work with point clouds, modify the deformation regularization functional, and modify how we solve the ordinary differential equations. This section studies the effect of all changes on the reconstruction performance by comparing RDA-INR to NDF on the liver dataset. 

We run NDF on the liver dataset, which we preprocess according to the NDF paper. We used the same training parameters the authors used except that we put the latent dimension to $32$ and increased the number of epochs to $3000$. For evaluating NDF, we scale the ground truth meshes, ground truth point clouds, and test reconstructions with a scale factor of $0.75$, which we also did on the training data for RDA-INR, as explained in Appendix \ref{app:datasets}. We copy the results of the RDA-INR model trained with Riemannian regularization from the main manuscript. 

Table \ref{tab:reconstruction_performance_RDA_INR_and_NDF} presents the reconstruction metrics on the test set of the liver dataset. RDA-INR and NDF show similar performance based on the Chamfer Distance. However, based on the Earth Mover distance, RDA-INR achieves better results. Hence, we conclude that the changes to NDF do not deteriorate its performance or even improve it when considering the Earth Mover distance.

\begin{table}[h!]
    \footnotesize
	\caption{\textbf{Reconstruction performance.} RDA-INR and NDF are evaluated by reconstructing the test set of the liver dataset. We calculate the average value and median (between brackets)of the Chamfer Distance (CD) and Earth Mover distance (EM). The CD values are of the order $10^{-4}$. The best (smallest) values are in bold.}
    
	\centering
	\begin{tabular}{c||rr|rr|}
		\toprule
		\multicolumn{1}{c@{\quad}}{\textbf{\textit{Model}}} & \multicolumn{4}{c}{\textbf{\textit{Metric}}}                 \\
		\cmidrule(r){1-1} \cmidrule(r){2-5} 
 & \multicolumn{2}{c|}{\textit{(CD)}} & \multicolumn{2}{c|}{\textit{(EM)}} \\
\cmidrule{1-5} \morecmidrules \cmidrule{1-5}
  \textit{NDF} & $5.61 $&$(4.97)$ & $0.0499  $&$(0.0479)$\\ 
  \textit{RDA-INR} & $\boldsymbol{5.31}$ & $(\boldsymbol{3.68})$ & $\boldsymbol{0.0385}$ & $(\boldsymbol{0.0340})  $ \\
		\bottomrule
	\end{tabular}
\label{tab:reconstruction_performance_RDA_INR_and_NDF}
\end{table}

\newpage

\section{Killing energy as standard LDDMM norm}\label{SM:kill_ener_as}
In the LDDMM literature, a commonly used norm is $\norm{\cdot}_V$ induced by $\langle \nu, \omega \rangle_V := \langle L\nu, \omega \rangle$ with $L : V \rightarrow V^*$, $\langle \cdot, \cdot \rangle$ the canonical duality pairing, and $\nu, \omega \in V$. In many papers, $L = (\text{id} - \alpha \Delta)^c$ for some $\alpha \in \mathbb{R}$ and $c \in \mathbb{N}$, and $\langle L\nu, \omega \rangle := \langle L\nu, \omega \rangle_{L^2(\Omega)}$. Alternatively, sometimes a slight adaptation of $L$ to $L = (\text{id} - \alpha \Delta - b\nabla \cdot \nabla^T)^c$ is used \cite{hinkle20124d, dupuis1998variational, christensen1996deformable, yang2017quicksilver}, where $\nabla \cdot \nabla^T \nu = \nabla \cdot (J\nu)^T$ with the divergence taken row-wise. In this section, we show that under certain conditions the Killing energy can be interpreted as $\norm{\nu}_V^2 = \langle L\nu, \nu \rangle_{L^2(\Omega)}$ with $L = (\text{id} -\alpha \Delta - b\nabla \cdot \nabla^T)$ for some $b$ and $\alpha$:
\begin{theorem}
Assume we use the following norm:
\begin{equation}
    \norm{\nu}_V^2 = \int_\Omega \frac{1}{2}\norm{J\nu + (J\nu)^T}_F^2 + \eta \norm{\nu}_2^2 \mathrm{d}x,
    \label{eq:thm_kill_ener_as_sob_kill_ener_norm}
\end{equation}
where $\Omega \subset \mathbb{R}^d$ is a bounded domain. Moreover, we assume that $\nu \in V$ is sufficiently many times differentiable and that:
\begin{equation}
    \int_{\partial \Omega} \langle \nu, (J\nu + (J\nu)^T) n \rangle_{l_2} \mathrm{d}x = 0,
    \label{eq:thm_kill_ener_as_sob_assumption}
\end{equation}
where $n$ is the normal to the boundary. Both assumptions are, for instance, satisfied when $V \subset C_0^2(\Omega, \mathbb{R}^d)$ where $C_0^2(\Omega, \mathbb{R}^d)$ is the space of twice continuously differentiable vector fields $\nu$ on the open and bounded domain $\Omega \subset \mathbb{R}^d$ such that both $\nu$ and its Jacobian $J\nu$ vanish on $\partial \Omega$. 

Given the assumptions, we have:
\begin{equation*}
    \norm{\nu}_V^2 = \int_\Omega \langle (\eta \,\normalfont{\text{id}} - \Delta - \nabla \cdot \nabla^T) \nu, \nu \rangle_{l_2}\mathrm{d}x = \int_\Omega \langle \Tilde{L} \nu,  \nu \rangle_{l_2}\mathrm{d}x,
\end{equation*}
where $\Tilde{L}:=\eta \,\normalfont{\text{id}} - \Delta - \nabla \cdot \nabla^T$. 
\label{thm:kill_energ_as_sob_norm}
\end{theorem}
\begin{proof}
First, define $\nu_i$ as the $i$-th component of the vector field $\nu$. As the Frobenius norm comes from an inner product and $\norm{A}_F^2 = \norm{A^T}_F^2$, we obtain
\begin{align*}
    \frac{1}{2}\norm{J\nu + (J\nu)^T}_F^2 & = \norm{J\nu}_F^2 + \langle J\nu, (J\nu)^T \rangle_F \\
    & = \sum_{i=1}^d \left\langle \nabla \nu_i, \nabla \nu_i \right \rangle_{l^2} + \left \langle \nabla \nu_i, \frac{\partial}{\partial x_i} \nu \right \rangle_{l_2} \\
    & = \sum_{i=1}^d \left \langle \nabla \nu_i, \nabla \nu_i + \frac{\partial}{\partial x_i} \nu \right \rangle_{l_2}.
\end{align*}
Using this identity, we obtain:
\begin{equation*}
    \int_\Omega \frac{1}{2}\norm{J\nu + (J\nu)^T}_F^2 \mathrm{d}x  = \int_{\Omega} \sum_{i=1}^d \left \langle \nabla \nu_i, \nabla \nu_i + \frac{\partial}{\partial x_i} \nu \right \rangle_{l_2}\mathrm{d}x = \sum_{i=1}^d \int_{\Omega} \left \langle \nabla \nu_i, \nabla \nu_i + \frac{\partial}{\partial x_i} \nu \right \rangle_{l_2}\mathrm{d}x.
\end{equation*}
Subsequently, using $\nabla \cdot (\nu_i \nabla \nu_i) = \left \langle \nabla \nu_i, \nabla \nu_i \right\rangle_{l^2} + \nu_i \Delta \nu_i$ and $\nabla \cdot (\nu_i \frac{\partial}{\partial x_i} \nu) = \left \langle \nabla \nu_i, \frac{\partial}{\partial x_i} \nu \right\rangle_{l^2} + \nu_i \nabla \cdot ( \frac{\partial}{\partial x_i} \nu)$, we get: 
\begin{equation*}
    \int_\Omega \frac{1}{2}\norm{J\nu + (J\nu)^T}_F^2 \mathrm{d}x = \sum_{i=1}^d \int_{\Omega} \nabla \cdot \left (\nu_i \left (\nabla \nu_i + \frac{\partial}{\partial x_i}\nu\right)\right) - \nu_i \left(\Delta \nu_i + \nabla \cdot \left ( \frac{\partial}{\partial x_i} \nu\right)\right)\mathrm{d}x .
\end{equation*}
Applying the divergence theorem yields:
\begin{align*}
    \int_\Omega \frac{1}{2}\norm{J\nu + (J\nu)^T}_F^2 \mathrm{d}x = \sum_{i=1}^d & \left( \int_{\partial \Omega} \left \langle \nu_i \left (\nabla \nu_i + \frac{\partial}{\partial x_i}\nu\right), n\right \rangle_{l_2} \mathrm{d}x - \right.\\
    & \, \, \, \, \left. \int_\Omega \nu_i \left(\Delta \nu_i + \nabla \cdot \left ( \frac{\partial}{\partial x_i} \nu\right)\right)\mathrm{d}x \right). 
\end{align*}
Doing some rewriting yields:
\begin{equation*}
    \int_\Omega \frac{1}{2}\norm{J\nu + (J\nu)^T}_F^2 \mathrm{d}x = \int_{\partial \Omega} \left \langle \nu, \left (J\nu + (J\nu)^T \right)n \right \rangle_{l^2} \mathrm{d}x - \int_{\Omega} \left \langle \nu, \left(\Delta + \nabla \cdot \nabla^T \right)\nu \right \rangle_{l^2} \mathrm{d}x.
\end{equation*}
Finally, using our assumption in Equation \eqref{eq:thm_kill_ener_as_sob_assumption} gives:
\begin{equation*}
    \int_\Omega \frac{1}{2}\norm{J\nu + (J\nu)^T}_F^2 \mathrm{d}x = - \int_{\Omega} \left \langle  \left(\Delta + \nabla \cdot \nabla^T \right)\nu, \nu \right \rangle_{l^2} \mathrm{d}x.
\end{equation*}
Using the above in combination with Equation \eqref{eq:thm_kill_ener_as_sob_kill_ener_norm}, we get:
\begin{align*}
        \norm{\nu}_V^2 & = \int_\Omega \frac{1}{2}\norm{J\nu + (J\nu)^T}_F^2 + \eta \norm{\nu}_2^2 \mathrm{d}x \\ 
        & = \int_\Omega \left \langle \left(\eta \,\normalfont{\text{id}}  - \Delta - \nabla \cdot \nabla^T\right)\nu, \nu \right \rangle_{l^2} \mathrm{d}x \\
        & = \int_\Omega \left \langle \Tilde{L}\nu, \nu \right \rangle_{l^2} \mathrm{d}x.
\end{align*}
\end{proof}
The theorem shows that our norm, together with the regularization constant $\sigma$, corresponds to the LDDMM norm with $L = (\text{id} - \alpha \Delta - b \nabla \cdot \nabla^T)^c$ for some $\alpha = b > 0$ and $c = 1$. Interestingly, in the few instances that LDDMM literature uses the $b > 0$ case, only Yang et al.\ \cite{yang2017quicksilver} use $c = 1$, while others typically use $c = 2$. While Yang et al.\ \cite{yang2017quicksilver} do not provide modeling for the $c = 1$ case, in the $c = 2$ case, the term with $b > 0$ stems from fluid mechanics modeling, with the identity mapping introduced to ensure that the operator is invertible.

Our approach differs in motivation by incorporating the Killing energy to induce a rigid body motion prior, which contrasts with the fluid mechanics perspective seen in other works. Therefore, our method, which results in the $c = 1$ case, is novel. Additionally, the $\nabla \cdot \nabla^T$ term in our regularizer can be viewed as a rigidity extension of the more common LDDMM operator $L = (\text{id} - \alpha \Delta)^c$ when using $c = 1$.

\newpage
\section{Implementation details} \label{app:impl_details}

\subsection{Parameterizing and solving the ordinary differential equation}
To solve optimization problems \eqref{eq:NN_optimization_problem} and \eqref{eq:NN_optimization_problem_IGR}, we need to parameterize the velocity vector field $v_\varphi$ and solve the resulting ordinary differential equation. Similarly to Gupta et al.\ \cite{gupta2020neural} and Sun et al.\ \cite{sun2022topology}, our model parameterizes $v_\varphi$ as a quasi-time-varying velocity vector field. Concretely, using $\chi_A$ as the indicator function of $A$, we define $K$ neural networks $v_{\varphi_k}:\Omega \times \mathbb{R}^{d_z} \rightarrow \mathbb{R}^d$ representing stationary velocity fields and define $v_\varphi$ as
\begin{equation}
    v_\varphi(x, t, z) = \sum_{k=1}^K \chi_{[\frac{k-1}{K}, \frac{k}{K})}(t) \cdot v_{\varphi_k}(x, z).
    \label{eq:v_varphi_quasi_time_varying_velocity_field}
\end{equation}
The main reason for this parameterization is that training time-varying velocity vector fields can be difficult as the model does not have training data for $0<t<1$. During training, we solve the ODE using an Euler discretization with $K$ time steps. Consequently, similar to ResNet-LDDMM \cite{amor2022resnet}, we approximate the ODE with a ResNet architecture.

\subsection{Pointwise loss} \label{app:pointwise_loss}
The models by \cite{zheng2021deep} and \cite{sun2022topology} are similar to our model. However, in contrast to our work, they do not use a Riemannian distance to regularize the time-dependent deformation of the template. Hence, we can not expect a physically plausible deformation that constitutes a geodesic. They use the pointwise regularization given by:
\begin{equation*}
    L_{pw} = \sum_{t \in T} \sum_{i=1}^N \sum_{j=1}^M \mathcal{L}_{0.25}\left(\norm{\phi_i(p_{ij}, t) - p_{ij}}_2\right),
\end{equation*}
where $\mathcal{L}_{0.25}$ is the Huber loss with parameter equal to $0.25$, $T$ is a set of predefined time instances at which to evaluate the pointwise loss, $\{p_{ij}\}_{j=1}^M$ are $M$ points from the domain, and $\phi_i(\cdot, t):=\phi_t^{z_i}(\cdot)$ with $\frac{\partial}{\partial t}\phi_t^z(x) = v_\varphi(\phi_t^z(x), t, z)$ and $\phi_0^z(x) = x$. The Huber loss is given by:
\begin{equation*}
    L_\delta(a) := \begin{cases}
        \frac{1}{2}a^2 & \mathrm{if} |a| \leq \delta \\
        \delta\cdot\left(|a| - \frac{1}{2}\delta \right) & \mathrm{otherwise}
    \end{cases}
\end{equation*}
We follow \cite{zheng2021deep} and \cite{sun2022topology} and choose $T=\{\lfloor K/4 \rfloor \cdot i \mid 1 \leq i \leq K / \lfloor 0.25 K \rfloor, i \in \mathbb{N} \}$ with $K$ the number of stationary velocity vector fields in $v_\varphi(x,t,z) = \sum_{k=1}^K \chi_{[(k-1)/K, k/K)}(t) \cdot v_{\varphi_k}(x, z)$.

In \cite{sun2022topology}, the pointwise loss is used to learn a template shape with similar features as the training shapes, which is one goal of our Riemannian regularization $\int_0^1 \norm{v(\cdot, t)}_V^2 \mathrm{d}t$ as well. Furthermore, note that when choosing $\norm{\cdot}_V = \norm{\cdot}_{L_2(\Omega)}$ and defining $\frac{\mathrm{d}}{\mathrm{d}t}x(t)=v_\varphi(x(t),t, z)$, our Riemannian regularization penalizes large differences between $x(1)$ and $x(0)$. As the pointwise loss also penalizes a large $\norm{x(1) - x(0)}_2$, it can be seen as a non-Riemannian version of our Riemannian LDDMM regularization.

\subsection{Neural network architectures}
\begin{figure}[t!]
     \centering
     \includegraphics[width=0.75\textwidth]{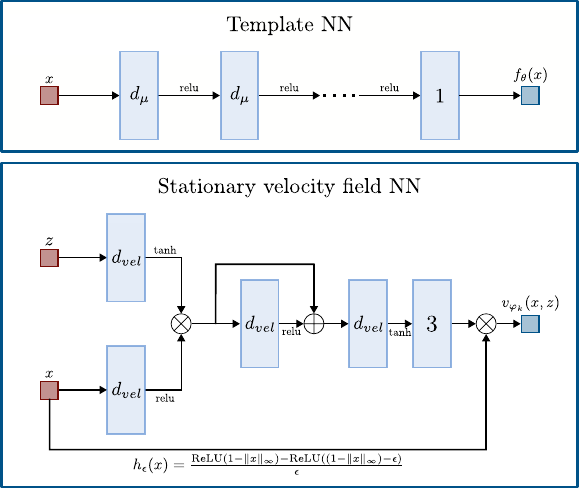}
     \caption{\textbf{Neural network architectures.} Architectures for the template neural network $f_\theta$ and the stationary velocity vector fields $v_{\varphi_k}$ in Equation \eqref{eq:v_varphi_quasi_time_varying_velocity_field}. The small red boxes indicate the latent code input $z$ and the spatial input $x$. Furthermore, the small blue box is an output, while the rectangles are linear layers with $d_{vel}$ or $d_\mu$ output dimensions. Finally, the $\oplus$ and $\otimes$ stand for elementwise addition and scalar/elementwise multiplication, respectively.}
    \label{fig:NN_architectures}
\end{figure}

Figure \ref{fig:NN_architectures} shows the used neural network architectures. For the template neural network, we use nearly the same architecture as the template neural network in \cite{deng2021deformed}. The only difference is that we use ReLU activation functions instead of sine activation functions. In our experiments, we clamp the output of the template neural network to the interval $[-0.5, 0.5]$. The stationary velocity vector field neural networks $v_{\varphi_k}$ in Equation \eqref{eq:v_varphi_quasi_time_varying_velocity_field} use the architecture in \cite{sun2022topology}. However, there are some differences. First, we add an extra linear layer at the output. We also add a scalar multiplication with the function $h_\epsilon$. This component ensures the velocity vector field is zero outside $\Omega$, which is needed to let the ODE be a diffeomorphism on $\Omega$. 

\subsection{Datasets} \label{app:datasets}
In our work, we use two datasets: a synthetic rectangles dataset and a shape liver dataset \cite{sun2022topology, chen2021deep}. 

The synthetic rectangles dataset is created by generating random boxes with edge lengths uniformly distributed in $[0.15, 0.85]$. Subsequently, these boxes are rotated using a random rotation matrix. For training the point cloud-based model, we sample 100 000 uniform points and corresponding normals from the meshes. For training using the occupancy data (see Appendix \ref{app:occnet_vs_IGR}), we uniformly sample 100 000 points from $\Omega=[-1,1]^3$ and calculate the signed distance to the boxes. Subsequently, we calculate the occupancy values from these signed distance values. The training dataset consists of 100 randomly generated parallelograms, while the test dataset consists of 20 parallelograms. 

For training the models on the liver dataset, we use the preprocessed data of \cite{sun2022topology}. The only additional preprocessing step is a scaling of their point cloud data and their mesh data. We multiply the points and the mesh vertices with a scaling factor of $0.75$ to make sure that all the livers are present in the unit cube. As training data, we sample 100 000 uniform points and corresponding normals from the meshes. Finally, we use the same train-test split as \cite{sun2022topology}, where the training dataset uses 145 samples and the test dataset uses 45 samples. 

The liver data has downsampled versions, which are used in Section \ref{sec:INR_for_stat_model}. These versions are obtained as outlined in the same section. For training, we use the same train-test split as for the original data. However, we exclude the samples with names contrast\_117, contrast\_136, and noncontrast\_034. The reason for this is that the original meshes of these samples were not watertight, which caused issues while generating downsampled meshes. However, since these problematic meshes were only in the training dataset, testing on the test set is not affected.

\subsection{Training details}
We jointly learn the latent codes $z_i$, the template implicit neural representation $f_\theta$, and the stationary velocity vector fields $v_{\varphi_k}$ ($k \in \{1, \ldots, K\}$). As the template neural network architecture $f_\theta$ comes from \cite{deng2021deformed} and the stationary velocity vector fields $v_{\varphi_k}$ from  \cite{sun2022topology}, we inherit the weight initialization schemes from these works. The latent codes $z_i \in \mathbb{R}^{d_z}$ are initialized by sampling from $\mathcal{N}(0, 1/d_z)$, as in \cite{sun2022topology}.

We use a batch size of 10 for each model and dataset pair. We approximate the expectations and spatial integrals in the point cloud data loss via Monte Carlo. For each expectation and integral, we sample 5000 points to estimate it. In particular, for the eikonal loss, we calculate the loss on 5000 random samples in $\Omega=[-1,1]^3$ and on 5000 warped surface samples. Subsequently, we average all the resulting values to obtain the final eikonal loss. Furthermore, for this eikonal loss, we only backpropagate the gradients concerning the parameters of the template INR $f_\theta$. Finally, for the experiments with the pointwise loss, the $p_{ij}$ values of the pointwise loss are taken as the 5000 surface samples as well as the 5000 random samples used to calculate the regularization corresponding to $\beta$ in Equation \eqref{eq:NN_optimization_problem_IGR}. 

For the experiment done in Appendix \ref{app:occnet_vs_IGR} with the occupancy data, we first sample 5000 points in $\Omega$ with their accompanying occupancy value. Half of the points lie inside, while the other half lie outside the shape. We calculate the binary cross entropy loss by averaging the binary cross entropies between the ground truth occupancy value and the estimated occupancy probability.

We update the neural network parameters and the latent codes $z_i$ via separate Adam optimizers for the latent codes, the template NN $f_\theta$, and velocity field NN $v_\varphi$. The initial learning rate for the latent codes is $10^{-3}$, for the $f_\theta$ parameters $5\cdot10^{-4}$, and for the $v_\varphi$ parameters $5\cdot10^{-4}$. Moreover, for each Adam optimizer, we use a learning rate scheduler that multiplies the learning rate with $0.7$ every 250 epochs. Finally, we bound the latent codes $z_i$ to be within the unit sphere. For the remainder of the hyperparameters, see Table \ref{tab:hyperparameters}.

Using the above considerations, we train our models on a high-performance computing cluster and use two NVIDIA A40 GPUs or two NVIDIA L40 GPUs. All of the GPUs have 48 GB of memory. The training took approximately 17 hours when using the A40's and approximately 13.5 hours when using the L40's. 

\subsection{Inference details}
For reconstructing shapes, we solve the following problem:
\begin{equation*}
    \min_z  \mathcal{D}_{rec}((\phi_1^z)^{-1} \cdot \mathcal{T}_\theta, \mathcal{S}) + \gamma \norm{z}_2^2,
\end{equation*}
where
\begin{itemize}
    \item $\gamma \in \mathbb{R}$,
    \item $\frac{\partial}{\partial t}\phi_t^z(x) = v_\varphi(\phi_t^z(x), t, z)$ and $\phi_0^z(x)=x$,
    \item $\mathcal{D}_{rec}((\phi_1^z)^{-1} \cdot \mathcal{T}_\theta, \mathcal{S}) = \expectation_{x\in \mathcal{S}} \left[ \lvert I(x, z, 1) \rvert \right]$,
    \item $I(x,z,t):=((\phi_t^z)^{-1} \cdot f_\theta)(x) := f_\theta(\phi_t^z(x))$.
\end{itemize}
We solve this optimization problem by running an Adam optimizer with an initial learning rate of $5\cdot10^{-2}$ for 800 iterations. At iteration 400 the learning rate is decreased by a factor of 10. Furthermore, we put $\gamma = 10^{-4}$ and we initialize the latent code from $\mathcal{N}(0, 0.01\cdot I)$.

\begin{table}[t!]
    \footnotesize
    \caption{\textbf{Hyperparameters for model training.} The hyperparameter values every trained model uses on a specific dataset (rectangles or liver). Here $N_h$ is the number of linear layers between the first and last linear layer in the template neural network (as presented in Figure \ref{fig:NN_architectures}). Furthermore, $c_{\text{PW}}$ denotes the regularization constant for the pointwise loss when training the model that uses this loss as the deformation regularizer.}
	\centering
	\begin{tabular}{c||c|c}
		\toprule
        \textbf{\textit{Hyperparameter}} & \textbf{\textit{Value (Rect.)}} & \textbf{\textit{Value (Liver)}} \\
		\cmidrule{1-3} \morecmidrules \cmidrule{1-3}
 Epochs & 4000 & 3000 \\
 \hline
 Latent dimension & 32 & 32\\ 
 \hline
 $d_{vel}$ & 512 & 512\\
  \hline
 $d_{\mu}$ & 256 & 256\\
 \hline
 $\epsilon$ & 0.05 & 0.05 \\
 \hline
 $K$ & 10 & 10 \\
 \hline 
 $d_\mu$ & 256 & 256 \\
 \hline
 $N_h$ & 5 & 5 \\
 \hline
 $\sigma^2$ & 0.025 & 0.002 \\
 \hline 
 $\tau$ & 0.01 & 0.01 \\
 \hline
 $\lambda$ & 0.005 & 0.005 \\
 \hline 
 $\beta$ & 1.5 & 1.5 \\
 \hline 
 $\alpha$ & 100 & 100 \\
 \hline
 $\eta$ & 0.05 & 50\\
 \hline
 $c_{\text{PW}}$ & 0.1 & 0.1 \\
		\bottomrule
	\end{tabular}
\label{tab:hyperparameters}
\end{table}

\subsection{DAE training and inference details}
To train and evaluate the DAE model \cite{bone2019learning}, we utilized the original code provided by the authors. For training DAE we tried to tune the hyperparameters to the best of our capabilities. 

First, we compute an initial template via Deformetrica's atlas estimation \cite{bone2018deformetrica}. We set the noise standard deviation of the current similarity metric to 0.01 and the kernel width to $0.1$. For the deformation kernel width, we used a value of $0.15$. Furthermore, we used ScipyLBFGS as the optimizer with an initial step size of $3\cdot10^{-5}$ and a very low convergence tolerance. Once we obtained the template, we employed Deformetrica's PGA function to obtain a $32$-dimensional latent space. We initialized the initial template with the estimated template and ran PGA with the same parameters as for the atlas estimation. 

We utilized the calculated PGA to initialize the DAE model. We use $15000$ epochs for this initialization. After that, we trained the DAE model for $15000$ epochs with a learning rate of $5\cdot10^{-4}$ and a batch size of $32$. We kept the template fixed as learning it resulted in a lot of instability. We chose a splatting grid size of $16$, a deformation grid size of $32$, and a kernel width of $0.15$ for the splatting layer and the current metric. We multiplied the data-fitting term by $10^4$ and the velocity field regularization term by $0.1$.

To evaluate the model, we used the encoder to estimate a latent vector. For the test data, we performed an additional optimization step for $1500$ iterations on the latent code using an Adam optimizer with a learning rate of $5\cdot10^{-3}$. This optimization step is the DAE+ version of the model, which was used to improve reconstruction performance on the test set. 

\end{document}